\documentclass[aps,twocolumn,nofootinbib,superscriptaddress,preprintnumbers,pra,10pt]{revtex4-1}

\usepackage{float}
\usepackage{arydshln}
\usepackage{colortbl}
\usepackage{amsmath,amssymb}
\usepackage{dsfont}
\usepackage{hyperref}
\usepackage{graphicx}
\usepackage{enumitem}
\usepackage{arydshln}
\usepackage{mathtools}
\usepackage{bbold}
\usepackage{soul}
\usepackage{slashed}
\usepackage{amsmath,bm}
\usepackage{yfonts}
\usepackage{lipsum}
\usepackage{slashed}
\usepackage{mathtools}
\usepackage{physics}

\makeatletter
\g@addto@macro\bfseries{\boldmath}
\makeatother

\newcommand{\be} {\begin{equation}}
\newcommand{\ee} {\end{equation}}
\newcommand{\bea} {\begin{eqnarray}}
\newcommand{\eea} {\end{eqnarray}}
\newcommand{\no} {\nonumber}
\newcommand{\cO}{{\mathcal O}}
\newcommand{\cL}{{\mathcal L}}

\newcommand{\cB}{{\mathcal B}}

\newcommand{\cC}{{\mathcal C}}
\renewcommand{\Re}{{\rm Re}}
\newcommand{\VV}{{\mathbf V}}
\newcommand{\Eps}{{\mathbf e}}
\newcommand{\DD}{{\mathbf  \Delta}}

\usepackage[normalem]{ulem}

\begin{document}

\preprint{ZU-TH 47/22}
\title{Confronting the vector leptoquark hypothesis with new low- and high-energy data}

\author{Jason Aebischer}
\author{Gino Isidori}
\author{Marko Pesut}
\author{Ben A. Stefanek}
\author{Felix Wilsch}

\affiliation{Physik-Institut, Universit\"at Z\"urich, CH-8057 Z\"urich, Switzerland}

\begin{abstract}
In light of new data we present an updated phenomenological analysis of the 
 simplified  $U_1$-leptoquark model addressing charged-current $B$-meson anomalies.
The analysis shows a good compatibility of low-energy data (dominated by the lepton flavor universality ratios 
$R_D$ and $R_{D^*}$) with the high-energy constraints posed by $pp\to \tau\bar\tau$ Drell-Yan data. 
We also show that present data are well compatible with a framework where 
the leptoquark couples with similar strength to both left- and right-handed third-generation fermions, a scenario that is well-motivated from a model building perspective.
We find that the high-energy implications of this setup will be probed at the 95\% confidence level in the high-luminosity phase of the LHC.
\end{abstract}

\maketitle
\allowdisplaybreaks

\section{Introduction}
The hypothesis of a vector leptoquark field ($U_1$), transforming as $(\mathbf{3},\mathbf{1},2/3)$ under the Standard Model (SM) 
gauge symmetry, with a mass in the TeV range has attracted intense interest in the last few years. 
At first, this interest arose from a purely phenomenological perspective, 
when it was realized that this field could offer a combined explanation of both the charged- 
and neutral-current $B$-meson anomalies~\cite{Alonso:2015sja,  Calibbi:2015kma, Barbieri:2015yvd, Bhattacharya:2016mcc}.
In fact, it was soon realized that the $U_1$ hypothesis is the only single-mediator 
explanation of the two sets of anomalies, while remaining well compatible with all available data~\cite{Buttazzo:2017ixm,Kumar:2018kmr,Angelescu:2018tyl}.
After these phenomenological analyses, a purely theoretical interest also began to grow with the realization that the $U_1$ hypothesis naturally points to an underlying $SU(4)$ Pati-Salam like~\cite{Pati:1974yy} symmetry unifying quarks and leptons~\cite{Barbieri:2015yvd}. In addition, the flavor structure of the $U_1$ couplings suggested by data hinted towards new dynamics potentially connected to the origin of the Yukawa
hierarchies~\cite{Barbieri:2015yvd,Buttazzo:2017ixm}.

These observations motivated an intense theoretical effort to build more complete models hosting a TeV-scale $U_1$ field.
Among them, a particularly compelling class is that of so-called ``4321" gauge models~\cite{DiLuzio:2017vat,Bordone:2017bld,Greljo:2018tuh,DiLuzio:2018zxy,Fuentes-Martin:2019ign,Fuentes-Martin:2020hvc}. In these models, the SM gauge symmetry is extended to $SU(4)_h\times SU(3)_l \times SU(2)_L \times U(1)_X$~\cite{DiLuzio:2017vat}, allowing the SM fermions to have flavor non-universal gauge charges~\cite{Bordone:2017bld},
such that the $U_1$ is coupled mainly to the heavy third-generation fermions.
It has also been proposed that the 4321 structure at the TeV scale, whose phenomenology has been analysed in detail in~\cite{Cornella:2019hct,Cornella:2021sby},
could be the first layer of a more ambitious multi-scale construction~\cite{Panico:2016ull,Bordone:2017bld,Allwicher:2020esa,Barbieri:2021wrc}.
This class of models are able to explain both the origin of the Yukawa hierarchies as well as stabilize the SM Higgs sector, as in~\cite{Fuentes-Martin:2020bnh,Fuentes-Martin:2020pww, Fuentes-Martin:2022xnb}. 
Alternative approaches to embed the $U_1$ in extended gauge groups and/or describe it in the context of composite models
have been proposed in~\cite{Assad:2017iib,Calibbi:2017qbu,Barbieri:2017tuq,Blanke:2018sro,Balaji:2018zna,Dolan:2020doe,Dolan:2020doe,King:2021jeo,FernandezNavarro:2022gst}, while additional recent phenomenological studies 
about the $U_1$ have been presented in~\cite{Angelescu:2021lln,Bhaskar:2021pml,Barbieri:2022ikw,Haisch:2022afh}.

Since the latest phenomenological studies, two sets of experimental data providing additional information about the 
leading $U_1$ couplings to third-generation fermions have appeared. On the low energy side,  LHCb has reported an updated measurement of the Lepton Flavor Universality (LFU) ratio $R_{D^*}$ and the first measurement of $R_D$ at a hadron collider~\cite{LHCbRd}, with the ratios defined as
\be
R_H = \Gamma(B \to H \tau\bar\nu)/\Gamma(B \to H \mu\bar\nu)\,.
\ee
On the high-energy side, 
new bounds on non-standard contributions to $\sigma(pp \to \tau\bar\tau)$ have been reported 
by CMS~\cite{CMS:2022goy,CMS:2022zks}.  
As pointed out first in~\cite{Faroughy:2016osc}, the $pp \to \tau\bar\tau$ process via $t$-channel $U_1$ exchange is a very sensitive probe of the $U_1$ 
couplings  to third-generation fermions, even for relatively high $U_1$ masses.
Interestingly enough, CMS data currently indicates a $3\sigma$ excess of events in $pp \to \tau\bar\tau$, well compatible 
with a possible $U_1$ contribution~\cite{CMS:2022zks}. However, no excess in $pp \to \tau\bar\tau$
is observed by ATLAS~\cite{ATLAS:2020zms} (although this analysis is not optimized for non-resonant $U_1$ contributions),  making drawing any conclusions about this excess premature.  
Still, these new data motivate a closer investigation about the compatibility of low- and high-energy observables under the  $U_1$ hypothesis,  which is the main goal of this paper.
We will pursue this goal in a general, bottom-up perspective by focusing only on the leading $U_1$ couplings to third-generation leptons while avoiding details that depend
on the specific ultraviolet (UV) completions of the model as much as possible. 

The paper is organized as follows: in Sec.~\ref{sec:model} we introduce the simplified model employed to analyze both low- and high-energy data. Particular attention is devoted to determine the (quark) flavor structure of the $U_1$ couplings, which is essential to relate the different amplitudes we are interested in ($b\to c\tau\bar\nu$ and 
$b\to u\tau\bar\nu$ at low energies, $b \bar b \to \tau\bar \tau$ at high energy). 
 In Sec.~\ref{sec:obs}, we perform a $\chi^2$-fit in our simplified model to determine the parameter space preferred by low-energy data. We then investigate the compatibility of the preferred low-energy parameter space with high-$p_T$ constraints from $pp \to \tau\bar\tau$. The  conclusions are summarised in Sec.~\ref{sec:conc}.
The Appendix~\ref{sect:appendix} contains a summary of the preferred 
parameter-space region in view of future searches.

\section{Model}
\label{sec:model}

The starting point of our analysis is the hypothesis of a massive $U_1$ field, coupled dominantly to third-generation fermions. 
Focusing on third-generation leptons, and assuming no leptoquark (LQ) couplings to
light right-handed fields (which are severely constrained by data, see e.g.~\cite{Fuentes-Martin:2019mun,Cornella:2019hct}),
we restrict our attention to the following terms in the LQ current:
\be
  J_U^\mu =\frac{g_U}{ \sqrt{2} } \left[\overline{q}^3_L\gamma^\mu \ell^3_L+\beta_R\,\overline{d}^3_R\gamma^\mu e^3_R+ 
  \sum_{k=1,2} \epsilon_{q_k} \, \overline{q}^k_L\gamma^\mu \ell^3_L\right]\,.
\label{eq:Jmu}
\ee
Here the right-handed fields and  the lepton doublet are understood to be in the corresponding mass-eigenstate basis, 
while the basis for the left-handed quarks is left generic and will be discussed in detail later on.

Integrating out the LQ field at the tree level leads to the effective interactions 
\bea
\label{eq:SMEFTLag}
\cL^{\rm LQ}_{\rm EFT} &=& -\frac{2}{v^2}  \Big[C_{LL}^{ij\alpha\beta}\, \cO^{ij\alpha\beta}_{LL}  +
C_{RR}^{ij\alpha\beta}\, \cO^{ij\alpha\beta}_{RR} \no\\  
&& \qquad +  \left(  C_{LR}^{ij\alpha\beta}\,
\cO^{ij\alpha\beta}_{LR}+{\rm h.c.} \right) \Big] ,
\eea
where
\begin{align}
\begin{aligned}
 \cO^{ij\alpha\beta}_{LL}  &= (\bar q_{L}^{\,i} \gamma_{\mu}  \ell_{L}^{\alpha}) (\bar \ell_{L}^{\beta} \gamma^{\mu}  q_{L}^{\,j} )\,, \no \\
 \cO^{ij\alpha\beta}_{LR} &=   (\bar q_{L}^{\,i} \gamma_{\mu}  \ell_{L}^{\alpha}) (\bar e_{R}^{\beta} \gamma^{\mu} d_{R}^{\,j} )\,,  \no \\
	 \cO^{ij\alpha\beta}_{RR} &= (\bar d_{R}^{\,i} \gamma_{\mu}  e_{R}^{\alpha}) (\bar e_{R}^{\beta} \gamma^{\mu}  d_{R}^{\,j})  \,.
\label{eq:opbasis}
\end{aligned}
\end{align}
The normalization factor in the effective Lagrangian is $v=(\sqrt{2}\,G_F)^{-1/2} \approx 246$~GeV.
We also introduce the effective scale $\Lambda_U = \sqrt{2} M_U/g_U$, such that 
\be
C^{33\tau\tau}_{LL} = \frac{ v^2 }{2  \Lambda_U^2 }\,.
\ee

If we were interested only in 
$b\to c\tau\bar\nu$ transitions, we would have restricted our attention to the  
coefficients $C^{cb\tau\tau}_{LL(LR)}$.\footnote{Here and in the rest of this section the
up- or down-type flavor indices referred to $q_L^i$ indicate the 
corresponding $SU(2)_L$ doublet in a given (up- or down-type)
mass eigenstate.} However, in order to also address the interplay with
$b\to u\tau\bar\nu$ transitions and,
most importantly,  high-energy constraints, we need to analyze the relation among the 
$C^{cb\tau\tau}_{LL(LR)}$ and 
coefficients involving different quark flavors.

\subsection{Quark flavor structure}
\label{sect:flavor}
The flavor basis defined by $J_U^\mu$ can be considered the interaction basis for the LQ field.
To address its relation to the mass-eigenstate basis of up (or down) quarks we need to 
write down and diagonalize the Yukawa couplings in this basis. 

As in~\cite{Barbieri:2015yvd}, we work under the assumption of an approximate  $U(2)_f^3 = U(2)_Q \times  U(2)_U \times U(2)_D$ 
symmetry acting on the light quark generations. In the limit of unbroken 
symmetry, the parameters $\epsilon_{q_k}$ in~(\ref{eq:Jmu}) should vanish 
and only third-generation quarks have non-zero Yukawa couplings. 
To describe a realistic spectrum, we proceed by introducing two sets of $U(2)_f^3$ breaking terms: 
\bea
  \Eps_q\,,~  \VV_{u}\,,~  \VV_{d}   &~\sim~& \mathbf{2_Q}\,,  \label{eq:VQ} \\
  \DD_{u}\,,~\DD_{d}  &~\sim~& \mathbf{{\bar 2}_{U(D)}}\times \mathbf{2_Q}\,,  \label{eq:Delta}
\eea
where $\Eps_q$ denotes the vector $\Eps^T_q = ( \epsilon_{q_1}, \epsilon_{q_2})$. 
The leading $\mathbf{2_Q}$ terms control  the heavy$\, \to\, $light mixing in the left-handed sector, 
whereas the subleading $\mathbf{{\bar 2}_{U(D)}}\times \mathbf{2_Q}$ terms 
are responsible for the light Yukawa couplings.  

The hypothesis of {\em minimal} $U(2)_f^3$ breaking, 
proposed in~\cite{Barbieri:2011ci,Barbieri:2012uh} and employed in previous phenomenological 
analysis (see e.g.~\cite{Barbieri:2015yvd,Buttazzo:2017ixm,Fuentes-Martin:2019mun}), corresponds to the assumption of a single $\mathbf{2_Q}$~spurion, or the alignment 
of the  three terms in~(\ref{eq:VQ}) in $U(2)_Q$ space. Motivated by model-building considerations~\cite{Fuentes-Martin:2022xnb,Crosas:2022quq} 
and recent data, we do not enforce this assumption in what follows.
In addition to the minimal case,  we will consider also the 
possibility of a (small) misalignment of the three leading $U(2)_Q$-breaking terms. 
We thus use the approximate $U(2)_f^3$ symmetry 
more as an organising principle to classify the 
flavor-violating couplings in the theory, rather than a strict ansatz on the underlying flavor structure.

Under these assumptions, 
the $3\times 3$ Yukawa couplings can be written as ($f=u,d$):
\begin{equation}
Y_f = y_{f_3} \left(\begin{array}{c|c}
 \DD_f &  \VV_f \\ \hline
 0 & 1
\end{array}\right)\,.
\end{equation}
Without loss of generality, the residual flavor symmetry allows us to choose a basis where both
$\DD_u$ and $\DD_d$ are real. In this basis, the latter are diagonalised by a real orthogonal matrix,
\be 
\Delta_f = O_f   \times  {\rm diag}\left( \frac{ y_{f_1} }{y_{f_3}} , \frac{ y_{f_2} }{y_{f_3}}  \right)\,, \quad
 O_f = \left(\begin{array}{cc} c_f & s_f \\ -s_f & c_f  \end{array}\right)\,,
 \ee
where $s_f =\sin\theta_f$ and  $c_f =\cos\theta_f$, and  $\VV_f$
are in general two complex vectors, $\VV^T_f = (V_{f_1},  V_{f_2})$.

The natural size of the different mixing terms  can be deduced 
by the perturbative diagonalisation of $Y_u$ and~$Y_d$. 
Introducing unitary matrices $L_f$, defined by 
\be
L_f  Y_f Y_f^\dagger L_f^\dagger = {\rm diag}( y_{f_1}, y_{f_2} , y_{f_3})\,,
\ee
it follows that 
\be 
L_f  \approx  \left(\begin{array}{c|c}
 O_f^T &  0 \\ \hline
 0 & 1
\end{array}\right)
\left(\begin{array}{c|c}
 1 &  -\VV_f \\ \hline
 \VV_f^\dagger & 1
\end{array}\right)\,.
\ee
Since the elements of the 
Cabibbo, Kobayashi, Maskawa (CKM) 
matrix are 
given by $V_{ij} = (L_u L_d^\dagger)_{ij}$, we deduce
\be
V_{u_2, d_2} = O(\lambda^2)\,, \quad V_{u_1, d_1}= O(\lambda^3)\,,
\label{eq:Nat-size}
\ee
where $\lambda =|V_{us}|  \approx 0.22$, and 
\be
 s_d-s_u  = \lambda + O(\lambda^3)\,.
 \label{eq:s_d}
 \ee
Assuming a common origin
of the leading $U(2)_Q$-breaking terms,  consistently with (\ref{eq:Nat-size}) 
it is natural to assume
\be
\epsilon_{q_2} = O(\lambda^2) \gg \epsilon_{q_1} ~.
\ee

Everything discussed so far follows from the initial choice of symmetry breaking terms, as well as
the requirement of reproducing the observed pattern of the quark Yukawa couplings. 
As we shall see, the non-observation of large deviations from the SM in ${\Delta F=2}$ transitions will impose further general constraints. This will allow us to pin down the precise relation between the Yukawa couplings and the LQ interaction basis.

\subsubsection*{Down-alignment of heavy$\,\to \,$light   mixing.} 

In any realistic UV completion of the effective model considered here, there are also currents
 $J_q^\mu = \overline{q}^3_L\gamma^\mu q^3_L$, associated to neutral mediators close in mass to the $U_1$ LQ.
As discussed in~\cite{Baker:2019sli}, this is an unavoidable 
consequence of the closure of the algebra associated to $J^\mu_U$. In particular, this conclusion holds no matter if the $U_1$ is realized as a gauge boson or as a composite state.
This fact implies that we also expect the effective interaction
\be
\Delta \cL^{4q} = O(1) \times  \frac{1}{\Lambda_U^2} (\overline{q}^3_L\gamma^\mu q^3_L)^2 \,.
\ee
The latter can spoil the tight bounds on $B_{s(d)}$--$\bar B_{s(d)}$  mixing 
unless the $V_{d_i}$ that control the off-diagonal entries of $L_d$
are about one order of magnitude smaller with respect to their natural 
size in Eq.~(\ref{eq:Nat-size}).\footnote{Precise bounds in 4321 
gauge models have been discussed in~\cite{Cornella:2021sby}.}  
The smallness of these parameters makes them irrelevant for any other observable, so in the following we simply set $\VV_d = 0$.
Under this assumption, the rotation matrices take the form
\be 
L_d  \approx  \left(\begin{array}{ccc}
 c_d &  - s_d &  0 \\ 
 s_d & c_d & 0 \\
 0 &  0& 1 
\end{array}\right)\,, \qquad  L_u = V \times L_d\,,
\label{eq:Ld}
\ee
and the only remaining free parameter in the Yukawa coupling is $s_u$ (or $s_d$), 
which control the orientation of $\VV_u$ in $U(2)_Q$ space 
relative to the CKM vector $(V_{ub}, V_{cb})$:\footnote{Note that without loss of generality we can change the (overall) phase of the fields 
such that $V_{u_2}$ is real and set the CKM matrix to its standard phase convention. }
\be
 \frac{ V_{u_1} }{ V_{u_2}} = s_u  +  \frac{V_{ub}}{V_{cb}} + O(\lambda^3)  
 \label{eq:su_cond}
\ee

At this point it is convenient to  re-write $J_U$ in the down-quark mass eigenstate basis by introducing the 
effective couplings $\beta_L^{ij}$ as in~\cite{Cornella:2019hct,Cornella:2021sby}:
\be
  J_U^\mu =\frac{g_U}{ \sqrt{2} } \left[ \sum_{q=b,s,d} \beta_L^{q\tau} \overline{q}_L\gamma^\mu \tau_L+\beta_R\,\overline{d}^3_R\gamma^\mu e^3_R \right]\,.
\label{eq:Jmu-down}
\ee
Using the expression of $L_d$ in Eq.~\eqref{eq:Ld} we get  $\beta_L^{b\tau} = 1$ and 
\bea
\beta_L^{s\tau} &=&  c_d \epsilon_{q_2} + s_d  \epsilon_{q_1}   = O(\lambda^2) \,,  \\
  \beta_L^{d\tau} &=&  c_d \epsilon_{q_1} -  s_d  \epsilon_{q_2}   = O(\lambda^3)\,.
\eea
Under the assumption of minimal $U(2)^3_f$ breaking, i.e.~assuming the two ${\mathbf 2_Q}$ spurions $\Eps_q$ and  $\VV_u$ are 
aligned in $U(2)_Q$ space, it is easy to check that 
\be
\left. \frac{ \beta_L^{d\tau}  }{   \beta_L^{s\tau} }  \right|_{\textrm{minimal}~U(2)^3_f} = \frac{V_{td}^* }{ V_{ts}^*}\,.
\label{eq:beta-min}
\ee
Therefore in the minimal case the value of the free parameter $s_u$ is irrelevant: it is 
absorbed into the definition of $\beta_L^{s\tau}$.

\subsubsection*{Non-minimal $U(2)_Q$ breaking with light-quark up alignment.} 

An interesting case worth considering from a model-building perspective
is the limit $\epsilon_{q_1} \to 0$, or the limit where the LQ field does not couple 
to the first generation (in a generic basis where the light-family mixing is real). 
This limit necessarily implies a non-minimal $U(2)_Q$ breaking, or a misalignment between 
$\Eps_q$ and $\VV_u$, as can be deduced by Eq.~(\ref{eq:su_cond}).\footnote{Setting
$\epsilon_{q_1}=0$ in a  
basis where the light-family mixing is real is equivalent to the statement that there is
no non-trivial CP-violating phase 
between $\Eps_q$ and $\Delta_{u,d}$.
This prevents reproducing the physical
phase in the CKM matrix using only these spurions. 
Indeed the (complex) relation~(\ref{eq:su_cond}) implies that the two components in $\VV_u$ 
have a different phase in the basis where 
$\DD_{u,d}$ are real.}
As we discuss below, in this limit we are phenomenologically led to assume a real $\epsilon_{q_2}$ as well as approximate up alignment in the light-quark sector (i.e.~$s_u\approx 0$), in order to evade 
the tight constraints from $ K$--$\bar K$ and $D$--$\bar D$ mixing. 

The ${\Delta F=2}$ constraints on the light-quark sector are 
more model dependent than those derived from ${\Delta B=2}$ transitions, since they 
depend on how the $U(2)_Q$ breaking is transferred from the LQ current
to the neutral currents. If the latter preserve a $U(2)_Q$ invariant structure, then there is no constraint coming from the light-quark sector. However, it is not obvious how to justify this from a model-building point of view.

In the most realistic scenarios, $U(2)_Q$ is broken also in the neutral-current sector
by terms proportional to appropriate insertions of $\Eps_q$. In this case, and assuming $\VV_d = 0$,
the severe constraint from CP-violation in $\bar K$--$K$  mixing can be satisfied assuming a real $\Eps_q$.
However, this is not enough to simultaneously protect CP-violation in $\bar D$--$D$ mixing. As pointed out recently in \cite{Crosas:2022quq} (see also~\cite{Barbieri:2022ikw}),
the latter forces us to choose $s_u \lesssim 0.1 \, \lambda$, i.e.~an approximate up alignment in the light-quark sector.

In the phenomenological limit $s_u=0$ and $\VV_d = 0$, the light-quark fields 
in the interaction basis can be identified as 
\be
 \left( \begin{array}{c} q_L^1 \\[2pt]  q_L^2 \end{array} \right) =
 \left( \begin{array}{cc}  V_{ud} & V_{us} \\[2pt]  V_{cd}  &  V_{cs} \end{array} \right)  
 \left( \begin{array}{c}  d_L \\[2pt]  s_L \end{array} \right) \approx  \left( \begin{array}{c}  u_L \\[2pt]  c_L \end{array} \right) \,,
\ee
while $q^3_L \equiv b_L$. The $\beta_L^{i\tau}$ become approximately diagonal in the up-quark mass basis 
and, setting $\epsilon_{q_1} \to 0$, we  get 
\be
\beta_L^{c\tau} =  \epsilon_{q_2}\,. \qquad  \beta_L^{u\tau} =  0~.
\label{eq:beta-up}
\ee

In the following we will investigate the relation between $b\to c$ and $ b\to u$ transitions 
either assuming the minimal-breaking relation (\ref{eq:beta-min}), or employing the ansatz~(\ref{eq:beta-up}).

\subsection{Charged currents in the mass-eigenstate basis}

Following the notation of~\cite{Cornella:2021sby}, 
we re-write the part of $\cL^{\rm LQ}_{\rm EFT}$ relevant to $b\to c\tau\bar\nu$ 
transitions as 
\begin{align}
   \cL_{b\to c}
  = - \frac{4 G_F}{\sqrt{2}} V_{cb} & \bigg[
    \Big( 1 + \cC_{LL}^{c} \Big)
    (\bar c_L \gamma_\mu b_L) (\bar\tau_L \gamma^\mu \nu_L)  \no\\
& \quad  - 2\,\cC_{LR}^{c} \,
    (\bar c_L b_R) (\bar\tau_R\,\nu_L) \bigg] \,,
    \label{eq:Cc}
\end{align} 
 and similarly for $b\to u\tau\bar\nu$. 
 The effective coefficients $\cC^{c,u}_{LL(LR)}$ defined above are related to the coefficients in (\ref{eq:SMEFTLag}) by
\be
  \cC_{LL(LR)}^c = \frac{ C_{LL(LR)}^{cb\tau\tau} }{V_{cb} }\,, \qquad   \cC_{LL(LR)}^u = \frac{ C_{LL(LR)}^{ub\tau\tau} }{V_{ub} }\,.
 \ee
 Using the $\beta^{ij}_{L}$ introduced in (\ref{eq:Jmu-down}), we get 
 \bea
  \cC_{LL}^c &=& C^{33\tau\tau}_{LL} \bigg( 1 +  \sum_{i=s,d} \frac{V_{ci}}{ V_{cb} }\beta_L^{i\tau}  \bigg) \equiv  C^{33\tau\tau}_{LL}\bigg( 1+ \frac{\epsilon_q}{ |V_{cb}| }\bigg)\,, \quad \no \\
 \cC_{LR}^c &=&  \beta^*_R \, \cC_{LL}^c\,,
 \label{eq:charmWCs}
\eea
where we defined the effective parameter 
$\epsilon_q$ to simplify the notation.
Concerning the $b\to u$ coefficients, 
assuming the minimal-breaking relation (\ref{eq:beta-min})
we get 
\be
  \cC_{LL(LR)}^u =   \cC_{LL(LR)}^c\,,
\ee
whereas the non-minimal ansatz~(\ref{eq:beta-up}) leads to
\be
  \cC_{LL(LR)}^u =  \frac{ \cC_{LL(LR)}^c  }{1 + \epsilon_q/ |V_{cb}| }\,.
\ee

\section{Observables}
\label{sec:obs}

\subsection{Low-energy}
The values of the effective couplings $\cC^c_{LL}$ and $\cC^c_{LR}$
can be fit at low energies using the experimental information on the LFU ratios $R_D$, $R_{D^*}$, and $R_{\Lambda_c}$. We have explicitly checked 
that other poorly measured observables, such as polarisation asymmetries 
in $b\to c\tau\bar\nu$ transitions or the loose bound on 
$\cB(B^-_c \to \tau \bar \nu)$~\cite{Alonso:2016oyd}, do not currently provide additional constraints.\footnote{Using the bound 
$\cB(B^-_c \to \tau \bar \nu) \leq 0.3$, derived in~\cite{Alonso:2016oyd}, we deduce $|\cC^{c}_{LR}| \leq 0.33$,
which has no influence on the fit.}

The LHCb collaboration recently reported a combined measurement of $R_D$ and $R_{D^*}$ based on the $\tau \rightarrow \mu \nu\nu$ decay of
$R_D = 0.441 \pm 0.060_{\rm stat} \pm 0.066_{\rm syst}$ and 
$R_{D^*}= 0.281 \pm 0.018_{\rm stat} \pm 0.02 4_{\rm syst}$ 
with correlation $\rho = -0.43$~\cite{LHCb:2023zxo} , as well as an $R_{D^*}$ only measurement based on hadronic $\tau$-decays with the value $R_{D^*}= 0.257 \pm 0.012_{\rm stat} \pm 0.018_{\rm syst}$~\cite{CERNRDstar}. Together, these measurements shift the world average of these ratios to~\cite{HFLAV:2019otj} 
\bea
R_{D^*}^{\rm exp} &=& 0.284 \pm 0.013_{\rm total}\,, \\
R_{D}^{\rm exp} &=& 0.356 \pm 0.029_{\rm total} \,, 
\eea
with correlation $\rho = -0.37$. We fit these results within our model 
using the approximate numerical formulae reported 
in~\cite{Cornella:2021sby}: 
\begin{align}
    \frac{R_D}{R_D^{\text{SM}}} = \,\, &  |1+\cC^c_{LL}|^2-3.00\, {\rm Re}\left[\left(1+\cC_{LL}^c\right) \cC^{c\,*}_{LR}\right] \nonumber \\
    &+4.12|\cC^c_{LR}|^2\,, \label{eq:RD} \\
    \frac{R_{D^*}}{R_{D^*}^{\text{SM}}} =\,\,  &  |1+\cC^c_{LL}|^2-0.24\, {\rm Re}\left[\left(1+\cC_{LL}^c\right) \cC^{c\,*}_{LR}\right] \nonumber \\ 
    &+0.16| \cC^c_{LR}|^2\,, 
\end{align}
where the Wilson coefficients are understood to be renomalized at the scale $\mu=m_b$.
As reference values for the  SM predictions we use the HFLAV averages~\cite{HFLAV:2019otj}:\footnote{More details about the SM predictions of $R_D$ and $R_{D^*}$ and their uncertainties can be found in~\cite{MILC:2015uhg,Na:2015kha,Bernlochner:2017jka,Gambino:2019sif,Bordone:2019vic,Martinelli:2021onb}}
\begin{align}
    R_{D}^{\text{SM}} = 0.298(4)\,, \hspace{10mm}
    R_{D^*}^{\text{SM}} = 0.254(5)\,.
\end{align}

Concerning $R_{\Lambda_c}$, we use the 
approximate formula provided in~\cite{Becirevic:2022bev}, that in 
our notation reads
\begin{align}
\frac{R_{\Lambda_c}}{R_{\Lambda_c}^{\rm SM}} =& \,\,
|1+\cC^c_{LL}|^2-1.01\,{\rm Re}\left[\cC^{c}_{LR}+\cC_{LL}^c \cC^{c\,*}_{LR}\right] \nonumber \\
&+1.34|\cC^c_{LR}|^2\,.
\label{eq:RLambda}
\end{align}
As inputs we use the recent LHCb result, $R^{\rm exp}_{\Lambda_c} = 0.242\pm 0.076$~\cite{LHCb:2022piu},  and the SM value $R_{\Lambda_c}^{\rm SM} = 0.333(13)$~\cite{Becirevic:2022bev}. 

In the case of $b\to u \tau \nu$ transitions, the only relevant constraint at present is provided by 
$\mathcal{B}(B^-_u \rightarrow \tau \bar{\nu})$. Here the numerical expression reads \cite{Fuentes-Martin:2019mun} :
\begin{equation}
   \frac{\mathcal{B}\left(B^{-}_u \rightarrow \tau \bar{\nu}\right)}{\mathcal{B}\left(B^{-}_u \rightarrow \tau \bar{\nu}\right)^{\mathrm{SM}}}=\left|1+\cC_{LL}^u-2\chi_u \, \cC_{LR}^u\right|^2 \, ,
   \label{eq:BsubU}
\end{equation}
where $\chi_u =m_{B^{+}}^2 /\left[m_\tau\left(m_b+m_u\right)\right] \approx 3.75 $. The data we use are 
$\mathcal{B}\left(B^{-}_u \rightarrow \tau \bar{\nu}\right)^{\mathrm{exp}}=1.09(24) \times 10^{-4}$~\cite{ParticleDataGroup:2022pth}
 and $\mathcal{B}\left(B^{-}_u \rightarrow \tau \bar{\nu}\right)^{\mathrm{SM}}=0.812(54) \times 10^{-4}$~\cite{Bona:2022zhn}.

\begin{figure}[t]
    \centering
    \includegraphics[width=0.95\linewidth]{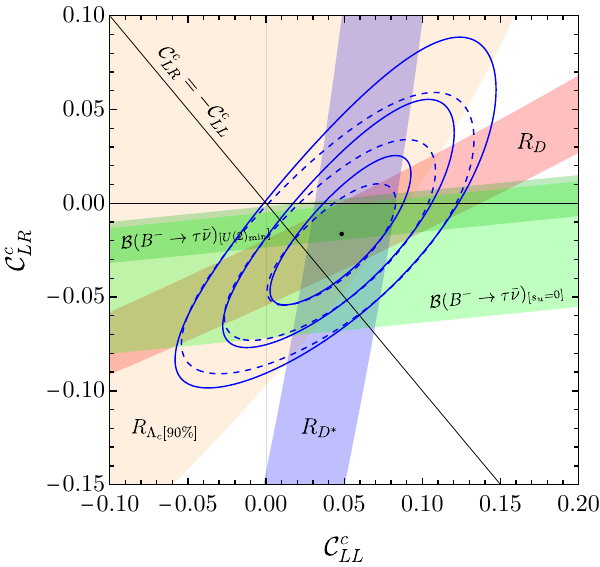}
    \caption{Determination of $\cC^c_{LL}$ and $\cC^c_{LR}$  from a $\chi^{2}$-fit to low-energy observables. 
    The Wilson coefficients, assumed to be real, are renormalized at the reference scale $\Lambda_{\rm UV}=1$~TeV.
    The blue ellipses denote the 1,~2, and 3$\sigma$ contours fitting only $b\to c$ observables. The black dot indicates the best fit point of $(0.05,-0.02)$. The dotted lines are obtained including 
    also $\mathcal{B}\left(B^{-}_u \rightarrow \tau \bar{\nu}\right)$ in the limit of up alignment. The $\Delta\chi^2 = 1$ regions preferred by each observable 
    are also indicated, except in the case of  $R_{\Lambda_c}$ where we give the 90\%~CL region (due to the large error).}
    \label{fig:lowE-fit}
\end{figure}

In Fig.~\ref{fig:lowE-fit} we report the best values of $\cC^c_{LL}$ and $\cC^c_{LR}$ as obtained from a $\chi^{2}$-fit to the low-energy observables.\footnote{As can be seen from Eqs.~(\ref{eq:RD}-\ref{eq:RLambda}), what matters for the low-energy fit in case of small Wilson coefficients is ${\rm Re}(\cC_{LR}^c)$, so in the fit we take $\cC_{LR}^c$ to be real for simplicity.} The values  reported in 
Fig.~\ref{fig:lowE-fit}  correspond to the Wilson coefficients renormalized at a reference high-scale $\Lambda_{\rm UV}=1$~TeV,
which is the most appropriate scale to compare low- and high-energy observables. Taking into account only the 
QCD-induced running, we set $\cC^c_{LL} (m_b) = \cC^c_{LL} (\Lambda_{\rm UV})$ and 
\be
\cC^c_{LR} (m_b) = \eta_S \, \cC^c_{LR} (\Lambda_{\rm UV})\,, \qquad  \eta_S\approx 1.6\,.
\ee

The first point to notice is that the SM point $(\cC^c_{LL}=\cC^c_{LR}=0)$ is excluded at the $3\sigma$ level.
The ${b\to c}$ observables  favor a region 
compatible with both a pure left-handed 
interaction ($\cC^c_{LR}=0$) as well as the case with equal magnitude right-handed currents $\cC^c_{LL}= -\cC^c_{LR}$. In both cases, the pull of the $U_1$ LQ hypothesis with respect to the SM is $\Delta\chi^2 = \chi^2_{\rm SM}-\chi^2_{\rm NP} \approx 11$, which is at the $3\sigma$ level. As first pointed out in~\cite{Bordone:2017bld}, 
the case where $\cC^c_{LL}= -\cC^c_{LR}$ is a natural benchmark
for a flavor non-universal gauge model, where  both left- and right-handed third-family quarks and leptons are unified in fundamental representations of $SU(4)$. As indicated by the dashed blue lines, the preferred region is  essentially unchanged if $\mathcal{B}(B^- \rightarrow \tau \bar{\nu})$ is added under the hypothesis of non-minimal $U(2)_Q$ breaking and up-alignment. In either case, we find a best fit point of $\cC^c_{LL}= 0.05$ and $\cC^c_{LR} = -0.02$. On the other hand,
the inclusion of $\mathcal{B}(B^- \rightarrow \tau \bar{\nu})$ 
under the hypothesis of minimal $U(2)_Q$ breaking
(dark green band) disfavors sizable right-handed currents.

\subsubsection*{Loop-induced contribution to $b\to s\ell\bar\ell$}

This analysis is focused on the leading couplings 
of the $U_1$ field to third-generation leptons. Hence, we do not discuss $b\to s\ell\ell$ transitions ($\ell=e,\mu$) in detail here.
However, we recall that the operator 
$\cO_{LL}^{sb\tau\tau}$ mixes via QED running~\cite{Aebischer:2017gaw}  into operators with light leptons
($\tau\bar\tau \to \ell\bar \ell$ loop). This results into a 
lepton-universal contribution to the $b\to s\ell\bar\ell$ 
Wilson coefficient $C_9$ \cite{Crivellin:2018yvo}, defined according to standard conventions (see e.g.~\cite{London:2021lfn,Altmannshofer:2021qrr}).
We will estimate the size of this effect using the results of the fit in Fig.~\ref{fig:lowE-fit}.

To this purpose, we note that besides the leading-log running from the
high-energy matching scale (i.e.~$M_U$) down to $m_b$, we should also include long distance (LD) contributions resulting from the one-loop matrix element of the semi-leptonic operator $\mathcal{O}^{sb\tau\tau}_{LL}$ \cite{Cornella:2020aoq}. Such contributions are analogous to the LD contributions from four-quark operators to the $b\to s\ell\bar\ell$
decay amplitude, which are present in the SM (see e.g.~\cite{Khodjamirian:2012rm}). 
The only difference is that the charm loop is replaced by a tau-lepton
loop. In full analogy to the factorizable part of the charm-loop contribution~\cite{Khodjamirian:2012rm}, 
also the (fully perturbative) LD tau-lepton contribution 
can be taken into account defining a $q^2$-dependent $C^{\rm eff}_9(q^2)$, where $q^2=m_{\ell\ell}^2$.
Considering also this effect, we find the following expression for the correction to $C^{\rm eff}_9$ induced by the $U_1$: 
\bea
 && 
 \Delta C_9^{\rm eff} (q^2=0) =
    \frac{C^{sb\tau\tau}_{LL}}{V_{ts}^* V_{tb}} \frac{2}{3}\left[\log{\left(\frac{M_U^2}{m^2_\tau}\right)}-1\right]\,,  \no \\
 &&  \quad =  -  \frac{ \cC^c_{LL}}{1+|V_{ts}|/\epsilon_q}
    \frac{2}{3}\left[\log{\left(\frac{M_U^2}{m^2_\tau}\right)}-1\right]\,.\quad 
\eea 
The last expression follows
from the relation between 
$C^{sb\tau\tau}_{LL}$ and $\cC^c_{LL}$, 
which can be 
deduced from Sect.~\ref{sect:flavor}.
For $\cC^c_{LL}=0.05$ (best fit point in Fig.~\ref{fig:lowE-fit}), $M_U=3$~TeV, and $\epsilon_q=2|V_{ts}|$, we get 
$\Delta C_9^{\rm eff} (0) \approx - 0.3$. 
While not solving all $b\to s\ell\bar\ell$
anomalies, such a correction leads to a significant 
improvement in the description of $b\to s\ell\bar\ell$ 
data~\cite{London:2021lfn,Altmannshofer:2021qrr,Cornella:2021sby}.

\subsection{High-energy}
Collider observables are known to provide rich information on the parameter space of vector leptoquark models~\cite{Faroughy:2016osc,Baker:2019sli,Angelescu:2021lln} explaining the $B$-meson anomalies, that is complementary to low-energy data~\cite{Allwicher:2022gkm,Cornella:2021sby}. A~variety of different underlying processes can be relevant at hadron colliders such as the LHC. The most important channels involving the $U_1$~leptoquark are:
\begin{itemize}
    \item Pair production $pp \to U_1^\ast U_1$,
    \item Quark-gluon scattering $qg \to U_1 \ell$,
    \item Quark-lepton fusion $q\ell \to U_1$,
    \item Drell-Yan $pp \to \ell\bar\ell$.
\end{itemize}
The main decay channels in models where the leptoquark predominantly couples to third generation fermions are $U_1 \to b\tau^+$ and~$U_1 \to t \bar{\nu}_\tau$. In the case of interest where $g_U \gtrsim g_s$, the Drell-Yan production channel due to $t$-channel LQ exchange provides the most stringent constraints on the parameter space.
Nevertheless, the other channels can still yield relevant information. For example, the searches for LQ pair production~\cite{Diaz:2017lit,Blumlein:1996qp,Dorsner:2018ynv} set a lower bound on the $U_1$~mass of $M_{U}\gtrsim 1.7\,\text{TeV}$~\cite{ATLAS:2021jyv,CMS:2020wzx}, which however only covers a small region of parameter space relevant for the explanation of the charged-current $B$-meson anomalies~\cite{Cornella:2021sby}. On the other hand, quark-gluon scattering~\cite{Dorsner:2018ynv,Hammett:2015sea,Mandal:2015vfa,Alves:2002tj} and resonant production through quark-lepton fusion~\cite{Haisch:2020xjd,Buonocore:2020erb,Greljo:2020tgv,Buonocore:2020nai,Buonocore:2022msy} will be important in case of a discovery, but they are not competitive at the moment.

Therefore, in the present analysis, we focus only on the non-resonant contributions of the $U_1$ vector LQ to Drell-Yan production. In particular, we are interested in the process $pp\to\tau\bar\tau$, with the main contribution due to $b\bar{b}\to\tau\bar\tau$, since we assume that the $U_1$ is predominantly coupled to third generation fermions. In such a scenario, the final state events are expected to contain an associated $b$-jet, due to gluon splitting $g \to b\bar{b}$ in the initial proton. 
We consider the CMS~\cite{CMS:2022goy} and ATLAS~\cite{ATLAS:2020zms} searches for the di-tau final state, based on the full LHC Run-II data sets. These searches provide results both in a $b$-tag channel, where an associated $b$-tagged jet is required in the final state, and in a $b$-veto channel, where the absence of any $b$-tagged jet is compulsory.


The contributions of the $U_1$~vector-leptoquark to Drell-Yan processes have recently been studied in Ref.~\cite{Haisch:2022afh} at next-to-leading order~(NLO) in QCD. Notice that in any UV completion the $U_1$~leptoquark is expected to be accommodated by further degrees of freedom with masses in the ballpark of the $U_1$~mass, that will lead to additional collider signatures~\cite{DiLuzio:2017vat,Greljo:2018tuh,Baker:2019sli,Cornella:2021sby}. These are, however, model dependent and thus not considered in the analysis at hand. Previous work investigating the connection of high-$p_T$ data with the low-energy observables for the 
 $B$-meson anomalies can be found in Refs.~\cite{Cornella:2021sby,Allwicher:2022gkm}. We extend these works by analysing the recent CMS di-tau search~\cite{CMS:2022goy} in addition to the already previously investigated ATLAS search~\cite{ATLAS:2020zms} for the same final state. Moreover, we use the results of Ref.~\cite{Haisch:2022afh} to extend the analysis incorporating NLO effects and to exploit the more constraining searches for di-tau final states in association with a $b$-jet.

\begin{figure}[t]
    \centering
    \includegraphics[height=0.95\linewidth]{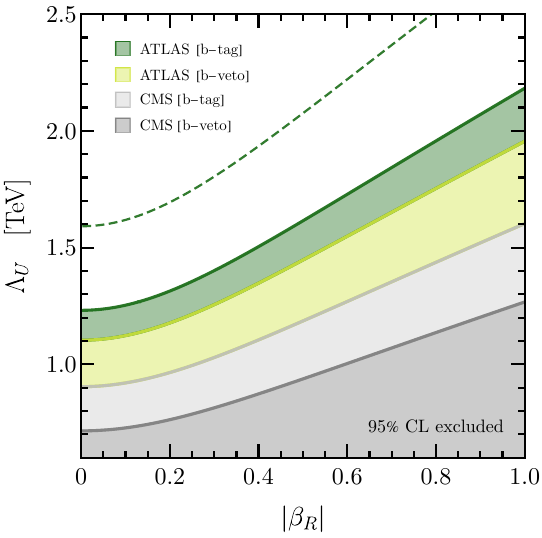}
    \caption{High-$p_T$ constraints on the $U_1$~model parameters~$\beta_R$ and~$\Lambda_U$ derived from the $pp\to\tau\bar\tau$ searches by CMS~\cite{CMS:2022goy}~(gray) and ATLAS~\cite{ATLAS:2020zms}~(green) in the $b$-tag and $b$-veto channels. The functional dependence is extracted using the \texttt{HighPT} package~\cite{Allwicher:2022mcg,Allwicher:2022gkm} and rescaled to the results presented in Ref.~\cite{Haisch:2022afh}. The shaded regions correspond to the excluded parameter space at~$95\,\%$~CL. The solid lines correspond to the constraints obtained using LHC run-II $(\sim 140\,\text{fb}^{-1})$ data, whereas the dashed line displays the projections for LHC's high luminosity phase~$(\sim 3\,\text{ab}^{-1})$ for the ATLAS $b$-tag search.}
    \label{fig:highpt-constraint}
\end{figure}

For our present study we use the \texttt{HighPT} package~\cite{Allwicher:2022mcg,Allwicher:2022gkm} to compute the $\chi^2$~likelihood of the EFT Lagrangian in Eq.~\eqref{eq:SMEFTLag} for the $b$-veto channel of the ATLAS di-tau search~\cite{ATLAS:2020zms}. We then rescale this result to match the NLO predictions derived in Ref.~\cite{Haisch:2022afh} for the $U_1$~leptoquark for the ATLAS~\cite{ATLAS:2020zms} and CMS~\cite{CMS:2022goy} searches in both $b$-tag and $b$-veto channels.\footnote{Ref.~\cite{Haisch:2022afh} also provides results for the CMS search~\cite{CMS:2022zks} for di-tau final states using angular observables. However, since such observables are currently not implemented in \texttt{HighPT}, we refrain from rescaling our likelihood obtained for the total-transverse mass~$m_{\tau\bar\tau}$ to this search.}

Minimizing the rescaled $\chi^2$~likelihoods with respect to the right-handed coupling~$\beta_R$ and the effective scale~$\Lambda_U$, we find the $95\,\%$~CL exclusion regions\footnote{The constraints presented in Fig.~\ref{fig:highpt-constraint} are obtained assuming $\epsilon_{q}=2|V_{ts}|$, but only exhibit a very mild dependence on~$\epsilon_{q_i}$.} shown in Fig.~\ref{fig:highpt-constraint}. 
The ATLAS di-tau search~\cite{ATLAS:2020zms}, shown in green, provides stronger exclusion limits than the corresponding CMS search~\cite{CMS:2022goy}, displayed in gray. This can be understood by noticing that a slight excess of events is observed in the high-$p_T$ tail in the latter search, weakening the constraints derived from it.
For both collaborations, the $b$-tag channels (dark green/light gray) yield more stringent constraints than the corresponding $b$-veto channels  (light green/dark gray), as anticipated.
As previously mentioned, this is because the signal comes dominantly from the process $b\bar{b}\to\tau\bar\tau$, where at least one bottom quark is likely to come from gluon splitting $(g\to b\bar{b})$ allowing to require an associated $b$-jet, which significantly reduces the background and thus yields stronger constraints.
Furthermore, it is evident that the scenarios with large right-handed currents~$\beta_R$ are tightly constrained by high-$p_T$ data. 


\begin{figure}[t]
    \centering
    \includegraphics[height=0.95\linewidth]{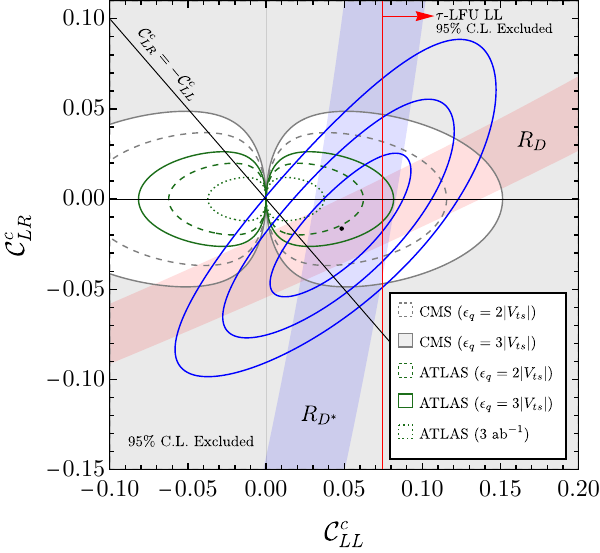}
    \caption{High-$p_T$ constraints superimposed on the low-energy fit. The red and blue bands represent the $\Delta\chi^2 = 1$~regions preferred by $R_{D}$ and~$R_{D^\ast}$. The blue lines correspond to the $1\sigma$, $2\sigma$, and $3\sigma$ contours of the combined low-energy fit including all $b\rightarrow c$ observables (dot = best fit point). 
    The high-$p_T$ exclusion limits derived from the $b$-tag channel of the CMS~\cite{CMS:2022goy} (ATLAS~\cite{ATLAS:2020zms}) search are given by regions outside of the gray (green) lines.
    On the other hand, the region inside the innermost dotted curve is our projection for the allowed parameter space from high-$p_T$ searches (in absence of a signal) with a luminosity of $3\,\text{ab}^{-1}$.  Finally, the region to the right of the red line is excluded by $\tau$-LFU tests assuming leading log running of $C^{33\tau\tau}_{LL}$. See text for more details.  }
    \label{fig:lowE-highpt}
\end{figure}

Next, we compare these high-$p_T$ results to the low-energy constraints derived in the previous section, by minimizing both likelihoods with respect to the Wilson coefficients~$\mathcal{C}_{LL}^c$ and~$\mathcal{C}_{LR}^c$, again evaluated at the reference high-scale $\Lambda_{\rm UV}=1$~TeV. The resulting fit is shown in Fig.~\ref{fig:lowE-highpt}, where the red and blue bands represent the preferred $\Delta\chi^2 = 1$~regions for the measurements of $R_{D}$ and~$R_{D^\ast}$. The blue lines correspond to the $1\sigma$, $2\sigma$, and $3\sigma$ contours of the combined low-energy fit including all $b\rightarrow c$ observables, whereas the gray (green) lines indicate the $95\,\%$~CL exclusion contours for the CMS (ATLAS) di-tau search using the $b$-tag channel.\footnote{Notice that the high-$p_T$ constraints are pinched at $\mathcal{C}_{LL}^c=0$ since this point corresponds to the limit $\beta_R\to\infty$ [see Eq.~\eqref{eq:charmWCs}].}  The solid and dashed lines correspond to the constraints obtained assuming $\epsilon_{q}=3|V_{ts}|$ and $\epsilon_{q}=2|V_{ts}|$, respectively.

As can be seen, the high-energy constraints are already very close to the parameter region favored by low-energy data.
To this purpose, it should be noted that scenarios with smaller $\epsilon_{q}$ are more constrained by high-$p_T$ as they require a lower scale $\Lambda_U$ to explain the charged-current anomalies (see Eq.~\eqref{eq:charmWCs}). On the other hand, values of $\epsilon_{q}$ larger than $3|V_{ts}|$ are both unnatural and 
highly disfavoured by $\Delta F=2$ constraints in UV complete models in the absence of fine-tuning. 

Due to the excess of events currently observed by CMS, the corresponding limits are significantly weaker than those of ATLAS. If interpreted as a signal, the CMS 
excess (which is further supported by a dedicated $t$-channel analysis~\cite{CMS:2022zks}) 
would favour the parameter region close to the CMS exclusion bounds in Fig.~\ref{fig:lowE-highpt}.
Given the low-energy constraints, this would 
in turn prefer a scenario with sizable right-handed couplings. 
On the other hand, ATLAS data are more compatible with low-energy data in the region of a pure left-handed coupling (though right-handed couplings remain viable).

Overall, the plot in Fig.~\ref{fig:lowE-highpt} shows that low- and high-energy data yield complementary constraints, 
and that a $U_1$ explanation of $R_{D^{(*)}}$ is  compatible with present $pp\to\tau\bar\tau$ data. 
This plot also shows that  future high-energy data will play an essential role in testing the $U_1$ explanation 
of charged-current $B$ anomalies. 
To illustrate this point, we indicate the projection for an integrated luminosity of~$3\,\text{ab}^{-1}$ by the shaded green central region in Fig.~\ref{fig:lowE-highpt}, which shows the potential of the high-luminosity phase of LHC assuming $\epsilon_{q}=2|V_{ts}|$. The projection was derived using the ATLAS $b$-tag search assuming that background uncertainties scale as the square-root of the luminosity. This projection shows that a large part of the relevant parameter space will be probed with the data sets expected from Run-III and the LHC high-luminosity phase. 

For completeness, in Fig.~\ref{fig:lowE-highpt}   we also indicate the region disfavoured by 
LFU tests in $\tau$ decays~\cite{Feruglio:2017rjo}: the region to the right of the red line is excluded 
by the experimental determination of $(g_{\tau}^W / g^{W}_{\mu,e})_{\ell,\pi,K}$~~\cite{HFLAV:2019otj}, using the leading-log (LL) running of $C^{33\tau\tau}_{LL} (1~{\rm TeV})$~\cite{Feruglio:2017rjo}, and 
setting $\epsilon_{q}=3|V_{ts}|$ (most conservative choice).
Due to their purely left-handed nature, $\tau$-LFU tests provide a strong constraint on the left-handed only hypothesis, potentially favouring scenarios with right-handed currents. However, this point comes with the caveat that additional contributions from new states in UV complete models can soften these bounds~\cite{Allwicher:2021ndi}.

\section{Conclusions}
\label{sec:conc}

In this paper we have analyzed the compatibility of the $U_1$ LQ explanation of the charged-current $B$-meson anomalies in light of new low- and high-energy data. 
To this purpose, we have first re-analysed in a
bottom-up and, to large extent, model-independent approach the assumptions necessary to relate the $U_1$ couplings appearing in $b\to c\tau\bar\nu$,
$b\to u\tau\bar\nu$, 
and $b \bar b \to \tau\bar \tau$ transitions.

Updating the fit to the  low-energy data, we find that the region preferred by $b\rightarrow c$ observables is equally compatible with a purely left-handed interaction, as well as with a scenario with right-handed currents of equal magnitude. 
The latter option is quite interesting, given sizable right-handed currents are a distinctive signature of models where the $U_1$ 
is embedded in a flavor non-universal gauge group~\cite{Bordone:2017bld}. In both cases, the pull of the $U_1$ hypothesis is at the $3\sigma$ level. The present low-energy fit 
already highlights the role of  $B_u \rightarrow \tau \bar \nu$ in pinning down the residual uncertainty on the flavor structure of the $U_1$ couplings. Indeed, this observable is expected to play an even more important role in the near future with the help of new data coming from Belle-II~\cite{Belle-II:2018jsg}.

Next, we examined collider constraints on the model, focusing on the $pp\to\tau\bar\tau$ Drell-Yan production channel mediated by $t$-channel $U_1$ exchange that provides the most stringent bounds. By superimposing these limits on the parameter space preferred by the low-energy fit, we conclude that constraints coming from the high-energy $pp\to\tau\bar\tau$ process are already closing in on the low-energy parameter space preferred by the charged-current $B$-meson anomalies. 

While low- and high-energy data are currently well compatible, a large fraction of the viable parameter space will be probed by the high-luminosity phase of the LHC. This is especially true in the case of equal magnitude left- and right-handed currents ($\cC^c_{LL}= -\cC^c_{LR}$), which has become more viable with the updated low-energy data and will be probed at the 95\% confidence level by the LHC. 
This will provide an exciting test of 
the well-motivated class of UV completions for the $U_1$ based 
on non-universal gauge groups, featuring 
quark-lepton unification for the third family  at the TeV scale~\cite{Bordone:2017bld,Greljo:2018tuh,Fuentes-Martin:2020bnh,Fuentes-Martin:2022xnb}.

\section*{Note Added}
While this project was under completion, an independent phenomenological analysis of charged-current $B$-meson anomalies,
including different leptoquark interpretations, has appeared~\cite{Iguro:2022yzr}.
Our results in Sect.~\ref{sec:obs} (low-energy fit) 
are compatible with those presented in~\cite{Iguro:2022yzr}.

\section*{Acknowledgements}
This project has received funding from the European Research Council (ERC) under the European Union's Horizon 2020 research and innovation programme under grant agreement 833280 (FLAY), and by the Swiss National Science Foundation (SNF) under contract 200020\_204428.

\appendix
\section{Preferred regions for $U_1$ couplings}
\label{sect:appendix}

In view of future searches of $U_1$ signals in channels involving $\tau$ leptons,
both at high and at low energies,  we provide here a summary of the preferred 
parameter-space region resulting from the 
low-energy fit performed in this paper.
We also report predictions for 
$\cB(B_s \to \tau^+\tau^-)$ and $\cB(B^+ \to K^+ \tau^+\tau^-)$, which can be considered the 
low-energy counterparts of $pp\to \tau\bar\tau$.

\begin{figure}
    \centering
    \includegraphics[width=0.9\linewidth]{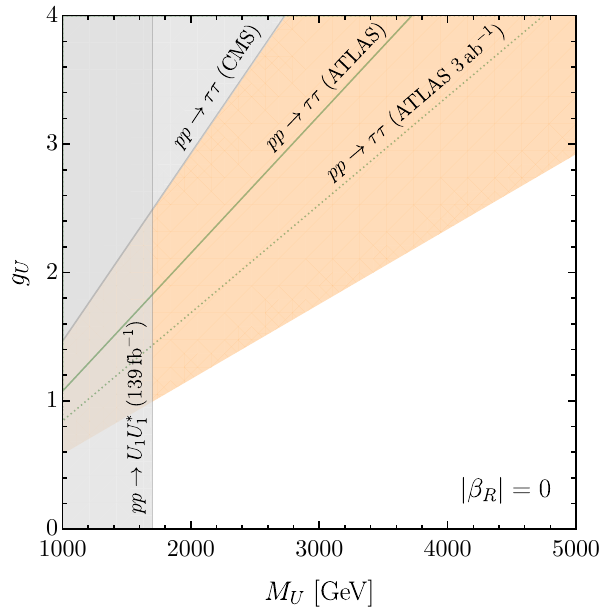}\\[5pt]
     \includegraphics[width=0.9\linewidth]{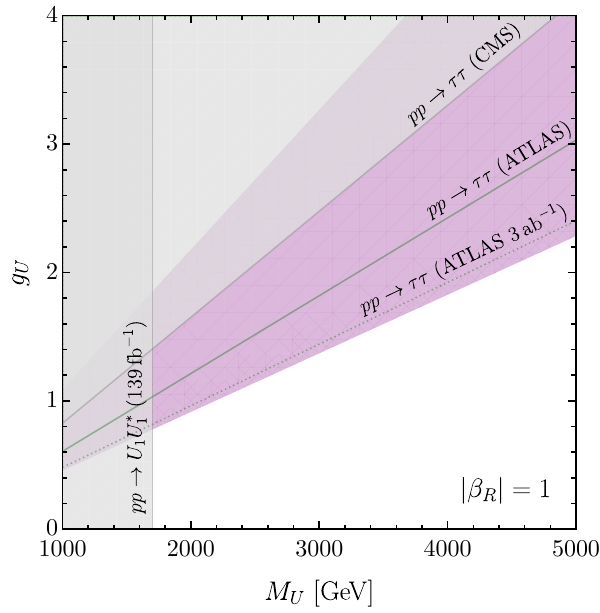}
    \caption{Preferred region at 90\% CL from low-energy charged-current data 
      for mass ($M_U$) and leading fermion coupling ($g_U$) of the  $U_1$ LQ.
      Top: Purely left-handed case ($\beta_R=0$). Bottom: 
      Pati-Salam-like case ($|\beta_R|=1$). The gray region and solid lines 
    indicate constraints of present high-energy searches at 95\% CL, while the dotted line gives the projected sensitivity at the HL-LHC with a luminosity of $3\,{\rm ab}^{-1}$. }
    \label{fig:gU_vs_mU}
\end{figure}

\begin{figure}
    \centering
    \includegraphics[width=0.95\linewidth]{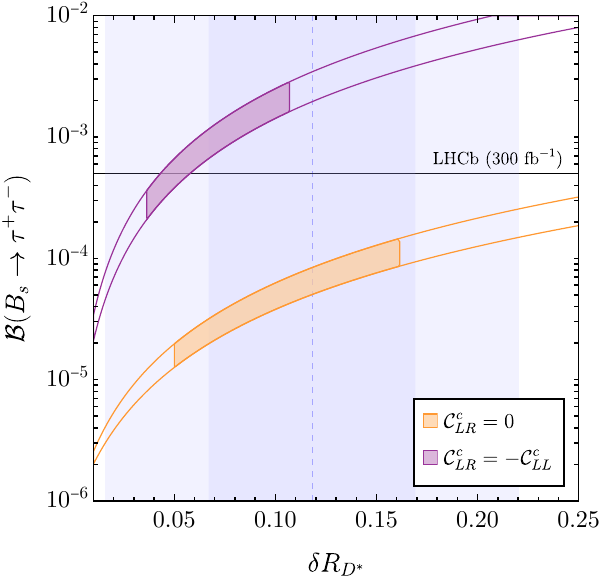} \\[5pt]
    \includegraphics[width=0.95\linewidth]{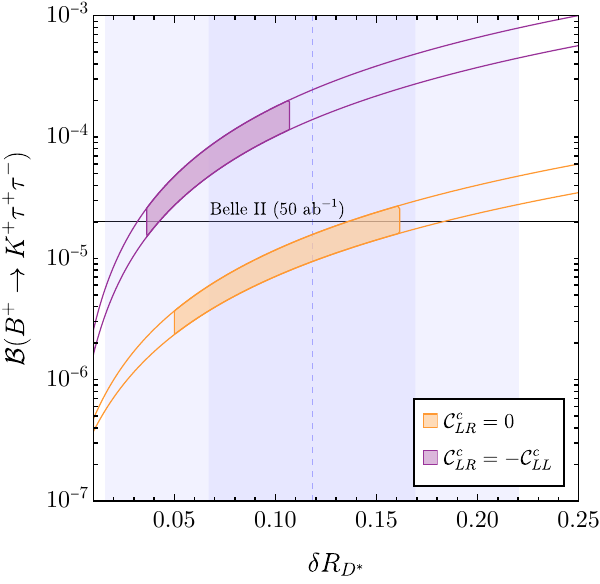}
    \caption{Predicted ranges for $\cB(B_s \rightarrow \tau^{+}\tau^{-})$ and $\cB(B^+ \to K^+ \tau^+\tau^-)$ as a function of $\delta R_{D^*}= R_{D^*}/R_{D^*}^{\text{SM}}-1$. The filled orange and purple colored regions correspond to the 
    90\% CL preferred regions from the low-energy charged-current fit. The blue vertical bands denote the present $1\sigma$ and $2\sigma$ experimental ranges for 
     $\delta R_{D^*}$.}
    \label{fig:bstautauPlot}
\end{figure}

The effective interaction between the $U_1$ field and fermion currents involving the 
$\tau$ lepton is $\mathcal{L}_{\rm int} = U_\mu J^{\mu}_U$,
with $J^{\mu}_U$ defined as in (\ref{eq:Jmu-down}).
By convention, we set $\beta_L^{b\tau} = 1$. 
The parameter $\beta_R$, which characterises different 
UV completions of the effective interaction with right-handed fermions,
should be treated as a free parameter.
In order to define precise benchmarks, we consider two reference cases for $\beta_R$:

\begin{enumerate}
\item $|\beta_R| = 0$ (Purely left-handed case) \\
Values preferred by low-energy data at 90\% CL:
\begin{align}
M_U/g_U &\in [0.69~{\rm TeV},\, 1.71~{\rm TeV}]\,,
\label{eq:guLH}
\end{align}
\item $|\beta_R| = 1$ (Pati-Salam-like LQ) \\
Values preferred by low-energy data at 90\% CL:
\begin{align}
M_U/g_U &\in [0.92~{\rm TeV},\, 2.19~{\rm TeV}]\,.
\label{eq:guRH}
\end{align}
\end{enumerate}

In Fig.~\ref{fig:gU_vs_mU} we show the present and future exclusion bounds in the $g_U$ vs.~$M_U$ plane from high-energy searches, as well as the region preferred by the low-energy fit corresponding to (\ref{eq:guLH}) and (\ref{eq:guRH}).

As discussed in the main text, low-energy data on charged currents alone are not able to provide a stringent constraint on $\beta_{s\tau}$. However, the latter is constrained by $\Delta M_{B_s}$ under general assumptions about the UV completion. The range we consider motivated in view of future experimental searches is
\begin{align}
\beta_{s\tau} \in [0.06,\, 0.16]\,,
\label{eq:Bsrange}
\end{align}
The $g_U/M_U$ ranges reported in (\ref{eq:guLH})
and (\ref{eq:guRH}) are obtained under this assumption, and setting $\beta_{d\tau}= 0$.

The theoretical expression for $B^+ \to K^+ \tau^+\tau^-$ reads~\cite{Cornella:2021sby}
\begin{align}
& 10^9\cB(B^+ \to K^+ \tau^+\tau^-) = 2.2 |\cC_9^{s\tau}|^2 + 6.0 |\cC_{10}^{s\tau}|^2 +8.3 |\cC_{S}^{s\tau}|^2 \nonumber \\
&+ 8.9 |\cC_{P}^{s\tau}|^2 + 4.8 {\rm Re}\left(\cC_{S}^{s\tau} \cC_{9}^{s\tau *} \right) + 5.9 {\rm Re}\left(\cC_{P}^{s\tau} \cC_{10}^{s\tau *} \right) \,,
\label{eq:BKtt}
\end{align}
where $\cC_{9}^{s\tau} = \cC_{\rm 9,SM} + \cC_{\rm 9,NP}^{s\tau}$ and  $\cC_{10}^{s\tau} = \cC_{\rm 10,SM} + \cC_{\rm 10,NP}^{s\tau}$, with $\cC_{\rm 9,SM} = 4.1$ and $\cC_{\rm 10,SM} = -4.2$.

In terms of the Wilson coefficients of $\cL^{\rm LQ}_{\rm EFT}$, 
the coefficients appearing in (\ref{eq:BKtt}) read
\begin{align}
\cC_{\rm 9,NP}^{s\tau} = -\cC_{\rm 10,NP}^{s\tau} &= -\frac{2\pi}{\alpha V_{ts}^{*}V_{tb}} \cC_{LL}^{sb\tau\tau} \,, \\
\cC_{S}^{s\tau} = -\cC_{P}^{s\tau} &= \frac{4\pi}{\alpha V_{ts}^{*}V_{tb}}\cC_{LR}^{sb\tau\tau} \,.
\end{align}

As far as $B_s \rightarrow \tau^{+}\tau^{-}$ is concerned, the  branching fraction 
can be decomposed as
\begin{align}
\frac{\mathcal{B}(B_s \rightarrow \tau^{+}\tau^{-})}{\mathcal{B}(B_s \rightarrow \tau^{+}\tau^{-})_{\rm SM}} = \bigg|1+\frac{\cC_{\rm 10,NP}^{s\tau}}{\cC_{\rm 10,SM}}+\frac{\chi_{s\tau}\cC_{P}^{s\tau}}{\cC_{\rm 10,SM}}\bigg|^2 & \nonumber \\
+ \left(1-\frac{4m_{\tau}^2}{m_{B_s}^2}\right)\bigg|\frac{\chi_{s\tau}\cC_{S}^{s\tau}}{\cC_{\rm 10,SM}}\bigg|^2  & \,,
\label{eq:Bstt}
\end{align}
where we have defined the chiral enhancement factor
\begin{equation}
\chi_{s\tau} = \frac{m_{B_s}^2}{2m_{\tau}(m_b+m_s)} \,.
\end{equation}

The model predictions for $\cB(B_s \rightarrow \tau^{+}\tau^{-})$ and $\cB(B^+ \to K^+ \tau^+\tau^-)$ corresponding to the two ranges in (\ref{eq:guLH})
and (\ref{eq:guRH}), as well as the flat range on $\beta_{s\tau}$ in (\ref{eq:Bsrange}),
are shown in Fig.~\ref{fig:bstautauPlot}.

%


\bibliography{main}

\begin{thebibliography}{83}%
\makeatletter
\providecommand \@ifxundefined [1]{%
 \@ifx{#1\undefined}
}%
\providecommand \@ifnum [1]{%
 \ifnum #1\expandafter \@firstoftwo
 \else \expandafter \@secondoftwo
 \fi
}%
\providecommand \@ifx [1]{%
 \ifx #1\expandafter \@firstoftwo
 \else \expandafter \@secondoftwo
 \fi
}%
\providecommand \natexlab [1]{#1}%
\providecommand \enquote  [1]{``#1''}%
\providecommand \bibnamefont  [1]{#1}%
\providecommand \bibfnamefont [1]{#1}%
\providecommand \citenamefont [1]{#1}%
\providecommand \href@noop [0]{\@secondoftwo}%
\providecommand \href [0]{\begingroup \@sanitize@url \@href}%
\providecommand \@href[1]{\@@startlink{#1}\@@href}%
\providecommand \@@href[1]{\endgroup#1\@@endlink}%
\providecommand \@sanitize@url [0]{\catcode `\\12\catcode `\$12\catcode
  `\&12\catcode `\#12\catcode `\^12\catcode `\_12\catcode `\%12\relax}%
\providecommand \@@startlink[1]{}%
\providecommand \@@endlink[0]{}%
\providecommand \url  [0]{\begingroup\@sanitize@url \@url }%
\providecommand \@url [1]{\endgroup\@href {#1}{\urlprefix }}%
\providecommand \urlprefix  [0]{URL }%
\providecommand \Eprint [0]{\href }%
\providecommand \doibase [0]{http://dx.doi.org/}%
\providecommand \selectlanguage [0]{\@gobble}%
\providecommand \bibinfo  [0]{\@secondoftwo}%
\providecommand \bibfield  [0]{\@secondoftwo}%
\providecommand \translation [1]{[#1]}%
\providecommand \BibitemOpen [0]{}%
\providecommand \bibitemStop [0]{}%
\providecommand \bibitemNoStop [0]{.\EOS\space}%
\providecommand \EOS [0]{\spacefactor3000\relax}%
\providecommand \BibitemShut  [1]{\csname bibitem#1\endcsname}%
\let\auto@bib@innerbib\@empty
\bibitem [{\citenamefont {Alonso}\ \emph {et~al.}(2015)\citenamefont {Alonso},
  \citenamefont {Grinstein},\ and\ \citenamefont
  {Martin~Camalich}}]{Alonso:2015sja}%
  \BibitemOpen
  \bibfield  {author} {\bibinfo {author} {\bibfnamefont {R.}~\bibnamefont
  {Alonso}}, \bibinfo {author} {\bibfnamefont {B.}~\bibnamefont {Grinstein}}, \
  and\ \bibinfo {author} {\bibfnamefont {J.}~\bibnamefont {Martin~Camalich}},\
  }\href {\doibase 10.1007/JHEP10(2015)184} {\bibfield  {journal} {\bibinfo
  {journal} {JHEP}\ }\textbf {\bibinfo {volume} {10}},\ \bibinfo {pages} {184}
  (\bibinfo {year} {2015})},\ \Eprint {http://arxiv.org/abs/1505.05164}
  {arXiv:1505.05164 [hep-ph]} \BibitemShut {NoStop}%
\bibitem [{\citenamefont {Calibbi}\ \emph {et~al.}(2015)\citenamefont
  {Calibbi}, \citenamefont {Crivellin},\ and\ \citenamefont
  {Ota}}]{Calibbi:2015kma}%
  \BibitemOpen
  \bibfield  {author} {\bibinfo {author} {\bibfnamefont {L.}~\bibnamefont
  {Calibbi}}, \bibinfo {author} {\bibfnamefont {A.}~\bibnamefont {Crivellin}},
  \ and\ \bibinfo {author} {\bibfnamefont {T.}~\bibnamefont {Ota}},\ }\href
  {\doibase 10.1103/PhysRevLett.115.181801} {\bibfield  {journal} {\bibinfo
  {journal} {Phys. Rev. Lett.}\ }\textbf {\bibinfo {volume} {115}},\ \bibinfo
  {pages} {181801} (\bibinfo {year} {2015})},\ \Eprint
  {http://arxiv.org/abs/1506.02661} {arXiv:1506.02661 [hep-ph]} \BibitemShut
  {NoStop}%
\bibitem [{\citenamefont {Barbieri}\ \emph {et~al.}(2016)\citenamefont
  {Barbieri}, \citenamefont {Isidori}, \citenamefont {Pattori},\ and\
  \citenamefont {Senia}}]{Barbieri:2015yvd}%
  \BibitemOpen
  \bibfield  {author} {\bibinfo {author} {\bibfnamefont {R.}~\bibnamefont
  {Barbieri}}, \bibinfo {author} {\bibfnamefont {G.}~\bibnamefont {Isidori}},
  \bibinfo {author} {\bibfnamefont {A.}~\bibnamefont {Pattori}}, \ and\
  \bibinfo {author} {\bibfnamefont {F.}~\bibnamefont {Senia}},\ }\href
  {\doibase 10.1140/epjc/s10052-016-3905-3} {\bibfield  {journal} {\bibinfo
  {journal} {Eur. Phys. J. C}\ }\textbf {\bibinfo {volume} {76}},\ \bibinfo
  {pages} {67} (\bibinfo {year} {2016})},\ \Eprint
  {http://arxiv.org/abs/1512.01560} {arXiv:1512.01560 [hep-ph]} \BibitemShut
  {NoStop}%
\bibitem [{\citenamefont {Bhattacharya}\ \emph {et~al.}(2017)\citenamefont
  {Bhattacharya}, \citenamefont {Datta}, \citenamefont {Gu\'evin},
  \citenamefont {London},\ and\ \citenamefont
  {Watanabe}}]{Bhattacharya:2016mcc}%
  \BibitemOpen
  \bibfield  {author} {\bibinfo {author} {\bibfnamefont {B.}~\bibnamefont
  {Bhattacharya}}, \bibinfo {author} {\bibfnamefont {A.}~\bibnamefont {Datta}},
  \bibinfo {author} {\bibfnamefont {J.-P.}\ \bibnamefont {Gu\'evin}}, \bibinfo
  {author} {\bibfnamefont {D.}~\bibnamefont {London}}, \ and\ \bibinfo {author}
  {\bibfnamefont {R.}~\bibnamefont {Watanabe}},\ }\href {\doibase
  10.1007/JHEP01(2017)015} {\bibfield  {journal} {\bibinfo  {journal} {JHEP}\
  }\textbf {\bibinfo {volume} {01}},\ \bibinfo {pages} {015} (\bibinfo {year}
  {2017})},\ \Eprint {http://arxiv.org/abs/1609.09078} {arXiv:1609.09078
  [hep-ph]} \BibitemShut {NoStop}%
\bibitem [{\citenamefont {Buttazzo}\ \emph {et~al.}(2017)\citenamefont
  {Buttazzo}, \citenamefont {Greljo}, \citenamefont {Isidori},\ and\
  \citenamefont {Marzocca}}]{Buttazzo:2017ixm}%
  \BibitemOpen
  \bibfield  {author} {\bibinfo {author} {\bibfnamefont {D.}~\bibnamefont
  {Buttazzo}}, \bibinfo {author} {\bibfnamefont {A.}~\bibnamefont {Greljo}},
  \bibinfo {author} {\bibfnamefont {G.}~\bibnamefont {Isidori}}, \ and\
  \bibinfo {author} {\bibfnamefont {D.}~\bibnamefont {Marzocca}},\ }\href
  {\doibase 10.1007/JHEP11(2017)044} {\bibfield  {journal} {\bibinfo  {journal}
  {JHEP}\ }\textbf {\bibinfo {volume} {11}},\ \bibinfo {pages} {044} (\bibinfo
  {year} {2017})},\ \Eprint {http://arxiv.org/abs/1706.07808} {arXiv:1706.07808
  [hep-ph]} \BibitemShut {NoStop}%
\bibitem [{\citenamefont {Kumar}\ \emph {et~al.}(2019)\citenamefont {Kumar},
  \citenamefont {London},\ and\ \citenamefont {Watanabe}}]{Kumar:2018kmr}%
  \BibitemOpen
  \bibfield  {author} {\bibinfo {author} {\bibfnamefont {J.}~\bibnamefont
  {Kumar}}, \bibinfo {author} {\bibfnamefont {D.}~\bibnamefont {London}}, \
  and\ \bibinfo {author} {\bibfnamefont {R.}~\bibnamefont {Watanabe}},\ }\href
  {\doibase 10.1103/PhysRevD.99.015007} {\bibfield  {journal} {\bibinfo
  {journal} {Phys. Rev. D}\ }\textbf {\bibinfo {volume} {99}},\ \bibinfo
  {pages} {015007} (\bibinfo {year} {2019})},\ \Eprint
  {http://arxiv.org/abs/1806.07403} {arXiv:1806.07403 [hep-ph]} \BibitemShut
  {NoStop}%
\bibitem [{\citenamefont {Angelescu}\ \emph {et~al.}(2018)\citenamefont
  {Angelescu}, \citenamefont {Be\v{c}irevi\'c}, \citenamefont {Faroughy},\ and\
  \citenamefont {Sumensari}}]{Angelescu:2018tyl}%
  \BibitemOpen
  \bibfield  {author} {\bibinfo {author} {\bibfnamefont {A.}~\bibnamefont
  {Angelescu}}, \bibinfo {author} {\bibfnamefont {D.}~\bibnamefont
  {Be\v{c}irevi\'c}}, \bibinfo {author} {\bibfnamefont {D.~A.}\ \bibnamefont
  {Faroughy}}, \ and\ \bibinfo {author} {\bibfnamefont {O.}~\bibnamefont
  {Sumensari}},\ }\href {\doibase 10.1007/JHEP10(2018)183} {\bibfield
  {journal} {\bibinfo  {journal} {JHEP}\ }\textbf {\bibinfo {volume} {10}},\
  \bibinfo {pages} {183} (\bibinfo {year} {2018})},\ \Eprint
  {http://arxiv.org/abs/1808.08179} {arXiv:1808.08179 [hep-ph]} \BibitemShut
  {NoStop}%
\bibitem [{\citenamefont {Pati}\ and\ \citenamefont
  {Salam}(1974)}]{Pati:1974yy}%
  \BibitemOpen
  \bibfield  {author} {\bibinfo {author} {\bibfnamefont {J.~C.}\ \bibnamefont
  {Pati}}\ and\ \bibinfo {author} {\bibfnamefont {A.}~\bibnamefont {Salam}},\
  }\href {\doibase 10.1103/PhysRevD.10.275} {\bibfield  {journal} {\bibinfo
  {journal} {Phys. Rev. D}\ }\textbf {\bibinfo {volume} {10}},\ \bibinfo
  {pages} {275} (\bibinfo {year} {1974})},\ \bibinfo {note} {[Erratum:
  Phys.Rev.D 11, 703--703 (1975)]}\BibitemShut {NoStop}%
\bibitem [{\citenamefont {Di~Luzio}\ \emph {et~al.}(2017)\citenamefont
  {Di~Luzio}, \citenamefont {Greljo},\ and\ \citenamefont
  {Nardecchia}}]{DiLuzio:2017vat}%
  \BibitemOpen
  \bibfield  {author} {\bibinfo {author} {\bibfnamefont {L.}~\bibnamefont
  {Di~Luzio}}, \bibinfo {author} {\bibfnamefont {A.}~\bibnamefont {Greljo}}, \
  and\ \bibinfo {author} {\bibfnamefont {M.}~\bibnamefont {Nardecchia}},\
  }\href {\doibase 10.1103/PhysRevD.96.115011} {\bibfield  {journal} {\bibinfo
  {journal} {Phys. Rev. D}\ }\textbf {\bibinfo {volume} {96}},\ \bibinfo
  {pages} {115011} (\bibinfo {year} {2017})},\ \Eprint
  {http://arxiv.org/abs/1708.08450} {arXiv:1708.08450 [hep-ph]} \BibitemShut
  {NoStop}%
\bibitem [{\citenamefont {Bordone}\ \emph {et~al.}(2018)\citenamefont
  {Bordone}, \citenamefont {Cornella}, \citenamefont {Fuentes-Martin},\ and\
  \citenamefont {Isidori}}]{Bordone:2017bld}%
  \BibitemOpen
  \bibfield  {author} {\bibinfo {author} {\bibfnamefont {M.}~\bibnamefont
  {Bordone}}, \bibinfo {author} {\bibfnamefont {C.}~\bibnamefont {Cornella}},
  \bibinfo {author} {\bibfnamefont {J.}~\bibnamefont {Fuentes-Martin}}, \ and\
  \bibinfo {author} {\bibfnamefont {G.}~\bibnamefont {Isidori}},\ }\href
  {\doibase 10.1016/j.physletb.2018.02.011} {\bibfield  {journal} {\bibinfo
  {journal} {Phys. Lett. B}\ }\textbf {\bibinfo {volume} {779}},\ \bibinfo
  {pages} {317} (\bibinfo {year} {2018})},\ \Eprint
  {http://arxiv.org/abs/1712.01368} {arXiv:1712.01368 [hep-ph]} \BibitemShut
  {NoStop}%
\bibitem [{\citenamefont {Greljo}\ and\ \citenamefont
  {Stefanek}(2018)}]{Greljo:2018tuh}%
  \BibitemOpen
  \bibfield  {author} {\bibinfo {author} {\bibfnamefont {A.}~\bibnamefont
  {Greljo}}\ and\ \bibinfo {author} {\bibfnamefont {B.~A.}\ \bibnamefont
  {Stefanek}},\ }\href {\doibase 10.1016/j.physletb.2018.05.033} {\bibfield
  {journal} {\bibinfo  {journal} {Phys. Lett. B}\ }\textbf {\bibinfo {volume}
  {782}},\ \bibinfo {pages} {131} (\bibinfo {year} {2018})},\ \Eprint
  {http://arxiv.org/abs/1802.04274} {arXiv:1802.04274 [hep-ph]} \BibitemShut
  {NoStop}%
\bibitem [{\citenamefont {Di~Luzio}\ \emph {et~al.}(2018)\citenamefont
  {Di~Luzio}, \citenamefont {Fuentes-Martin}, \citenamefont {Greljo},
  \citenamefont {Nardecchia},\ and\ \citenamefont {Renner}}]{DiLuzio:2018zxy}%
  \BibitemOpen
  \bibfield  {author} {\bibinfo {author} {\bibfnamefont {L.}~\bibnamefont
  {Di~Luzio}}, \bibinfo {author} {\bibfnamefont {J.}~\bibnamefont
  {Fuentes-Martin}}, \bibinfo {author} {\bibfnamefont {A.}~\bibnamefont
  {Greljo}}, \bibinfo {author} {\bibfnamefont {M.}~\bibnamefont {Nardecchia}},
  \ and\ \bibinfo {author} {\bibfnamefont {S.}~\bibnamefont {Renner}},\ }\href
  {\doibase 10.1007/JHEP11(2018)081} {\bibfield  {journal} {\bibinfo  {journal}
  {JHEP}\ }\textbf {\bibinfo {volume} {11}},\ \bibinfo {pages} {081} (\bibinfo
  {year} {2018})},\ \Eprint {http://arxiv.org/abs/1808.00942} {arXiv:1808.00942
  [hep-ph]} \BibitemShut {NoStop}%
\bibitem [{\citenamefont {Fuentes-Mart\'\i{}n}\ \emph
  {et~al.}(2020{\natexlab{a}})\citenamefont {Fuentes-Mart\'\i{}n},
  \citenamefont {Isidori}, \citenamefont {K\"onig},\ and\ \citenamefont
  {Selimovi\'c}}]{Fuentes-Martin:2019ign}%
  \BibitemOpen
  \bibfield  {author} {\bibinfo {author} {\bibfnamefont {J.}~\bibnamefont
  {Fuentes-Mart\'\i{}n}}, \bibinfo {author} {\bibfnamefont {G.}~\bibnamefont
  {Isidori}}, \bibinfo {author} {\bibfnamefont {M.}~\bibnamefont {K\"onig}}, \
  and\ \bibinfo {author} {\bibfnamefont {N.}~\bibnamefont {Selimovi\'c}},\
  }\href {\doibase 10.1103/PhysRevD.101.035024} {\bibfield  {journal} {\bibinfo
   {journal} {Phys. Rev. D}\ }\textbf {\bibinfo {volume} {101}},\ \bibinfo
  {pages} {035024} (\bibinfo {year} {2020}{\natexlab{a}})},\ \Eprint
  {http://arxiv.org/abs/1910.13474} {arXiv:1910.13474 [hep-ph]} \BibitemShut
  {NoStop}%
\bibitem [{\citenamefont {Fuentes-Mart\'\i{}n}\ \emph
  {et~al.}(2020{\natexlab{b}})\citenamefont {Fuentes-Mart\'\i{}n},
  \citenamefont {Isidori}, \citenamefont {K\"onig},\ and\ \citenamefont
  {Selimovi\'c}}]{Fuentes-Martin:2020hvc}%
  \BibitemOpen
  \bibfield  {author} {\bibinfo {author} {\bibfnamefont {J.}~\bibnamefont
  {Fuentes-Mart\'\i{}n}}, \bibinfo {author} {\bibfnamefont {G.}~\bibnamefont
  {Isidori}}, \bibinfo {author} {\bibfnamefont {M.}~\bibnamefont {K\"onig}}, \
  and\ \bibinfo {author} {\bibfnamefont {N.}~\bibnamefont {Selimovi\'c}},\
  }\href {\doibase 10.1103/PhysRevD.102.115015} {\bibfield  {journal} {\bibinfo
   {journal} {Phys. Rev. D}\ }\textbf {\bibinfo {volume} {102}},\ \bibinfo
  {pages} {115015} (\bibinfo {year} {2020}{\natexlab{b}})},\ \Eprint
  {http://arxiv.org/abs/2009.11296} {arXiv:2009.11296 [hep-ph]} \BibitemShut
  {NoStop}%
\bibitem [{\citenamefont {Cornella}\ \emph {et~al.}(2019)\citenamefont
  {Cornella}, \citenamefont {Fuentes-Martin},\ and\ \citenamefont
  {Isidori}}]{Cornella:2019hct}%
  \BibitemOpen
  \bibfield  {author} {\bibinfo {author} {\bibfnamefont {C.}~\bibnamefont
  {Cornella}}, \bibinfo {author} {\bibfnamefont {J.}~\bibnamefont
  {Fuentes-Martin}}, \ and\ \bibinfo {author} {\bibfnamefont {G.}~\bibnamefont
  {Isidori}},\ }\href {\doibase 10.1007/JHEP07(2019)168} {\bibfield  {journal}
  {\bibinfo  {journal} {JHEP}\ }\textbf {\bibinfo {volume} {07}},\ \bibinfo
  {pages} {168} (\bibinfo {year} {2019})},\ \Eprint
  {http://arxiv.org/abs/1903.11517} {arXiv:1903.11517 [hep-ph]} \BibitemShut
  {NoStop}%
\bibitem [{\citenamefont {Cornella}\ \emph {et~al.}(2021)\citenamefont
  {Cornella}, \citenamefont {Faroughy}, \citenamefont {Fuentes-Martin},
  \citenamefont {Isidori},\ and\ \citenamefont {Neubert}}]{Cornella:2021sby}%
  \BibitemOpen
  \bibfield  {author} {\bibinfo {author} {\bibfnamefont {C.}~\bibnamefont
  {Cornella}}, \bibinfo {author} {\bibfnamefont {D.~A.}\ \bibnamefont
  {Faroughy}}, \bibinfo {author} {\bibfnamefont {J.}~\bibnamefont
  {Fuentes-Martin}}, \bibinfo {author} {\bibfnamefont {G.}~\bibnamefont
  {Isidori}}, \ and\ \bibinfo {author} {\bibfnamefont {M.}~\bibnamefont
  {Neubert}},\ }\href {\doibase 10.1007/JHEP08(2021)050} {\bibfield  {journal}
  {\bibinfo  {journal} {JHEP}\ }\textbf {\bibinfo {volume} {08}},\ \bibinfo
  {pages} {050} (\bibinfo {year} {2021})},\ \Eprint
  {http://arxiv.org/abs/2103.16558} {arXiv:2103.16558 [hep-ph]} \BibitemShut
  {NoStop}%
\bibitem [{\citenamefont {Panico}\ and\ \citenamefont
  {Pomarol}(2016)}]{Panico:2016ull}%
  \BibitemOpen
  \bibfield  {author} {\bibinfo {author} {\bibfnamefont {G.}~\bibnamefont
  {Panico}}\ and\ \bibinfo {author} {\bibfnamefont {A.}~\bibnamefont
  {Pomarol}},\ }\href {\doibase 10.1007/JHEP07(2016)097} {\bibfield  {journal}
  {\bibinfo  {journal} {JHEP}\ }\textbf {\bibinfo {volume} {07}},\ \bibinfo
  {pages} {097} (\bibinfo {year} {2016})},\ \Eprint
  {http://arxiv.org/abs/1603.06609} {arXiv:1603.06609 [hep-ph]} \BibitemShut
  {NoStop}%
\bibitem [{\citenamefont {Allwicher}\ \emph {et~al.}(2021)\citenamefont
  {Allwicher}, \citenamefont {Isidori},\ and\ \citenamefont
  {Thomsen}}]{Allwicher:2020esa}%
  \BibitemOpen
  \bibfield  {author} {\bibinfo {author} {\bibfnamefont {L.}~\bibnamefont
  {Allwicher}}, \bibinfo {author} {\bibfnamefont {G.}~\bibnamefont {Isidori}},
  \ and\ \bibinfo {author} {\bibfnamefont {A.~E.}\ \bibnamefont {Thomsen}},\
  }\href {\doibase 10.1007/JHEP01(2021)191} {\bibfield  {journal} {\bibinfo
  {journal} {JHEP}\ }\textbf {\bibinfo {volume} {01}},\ \bibinfo {pages} {191}
  (\bibinfo {year} {2021})},\ \Eprint {http://arxiv.org/abs/2011.01946}
  {arXiv:2011.01946 [hep-ph]} \BibitemShut {NoStop}%
\bibitem [{\citenamefont {Barbieri}(2021)}]{Barbieri:2021wrc}%
  \BibitemOpen
  \bibfield  {author} {\bibinfo {author} {\bibfnamefont {R.}~\bibnamefont
  {Barbieri}},\ }\href {\doibase 10.5506/APhysPolB.52.789} {\bibfield
  {journal} {\bibinfo  {journal} {Acta Phys. Polon. B}\ }\textbf {\bibinfo
  {volume} {52}},\ \bibinfo {pages} {789} (\bibinfo {year} {2021})},\ \Eprint
  {http://arxiv.org/abs/2103.15635} {arXiv:2103.15635 [hep-ph]} \BibitemShut
  {NoStop}%
\bibitem [{\citenamefont {Fuentes-Mart\'\i{}n}\ and\ \citenamefont
  {Stangl}(2020)}]{Fuentes-Martin:2020bnh}%
  \BibitemOpen
  \bibfield  {author} {\bibinfo {author} {\bibfnamefont {J.}~\bibnamefont
  {Fuentes-Mart\'\i{}n}}\ and\ \bibinfo {author} {\bibfnamefont
  {P.}~\bibnamefont {Stangl}},\ }\href {\doibase
  10.1016/j.physletb.2020.135953} {\bibfield  {journal} {\bibinfo  {journal}
  {Phys. Lett. B}\ }\textbf {\bibinfo {volume} {811}},\ \bibinfo {pages}
  {135953} (\bibinfo {year} {2020})},\ \Eprint
  {http://arxiv.org/abs/2004.11376} {arXiv:2004.11376 [hep-ph]} \BibitemShut
  {NoStop}%
\bibitem [{\citenamefont {Fuentes-Martin}\ \emph {et~al.}(2021)\citenamefont
  {Fuentes-Martin}, \citenamefont {Isidori}, \citenamefont {Pag\`es},\ and\
  \citenamefont {Stefanek}}]{Fuentes-Martin:2020pww}%
  \BibitemOpen
  \bibfield  {author} {\bibinfo {author} {\bibfnamefont {J.}~\bibnamefont
  {Fuentes-Martin}}, \bibinfo {author} {\bibfnamefont {G.}~\bibnamefont
  {Isidori}}, \bibinfo {author} {\bibfnamefont {J.}~\bibnamefont {Pag\`es}}, \
  and\ \bibinfo {author} {\bibfnamefont {B.~A.}\ \bibnamefont {Stefanek}},\
  }\href {\doibase 10.1016/j.physletb.2021.136484} {\bibfield  {journal}
  {\bibinfo  {journal} {Phys. Lett. B}\ }\textbf {\bibinfo {volume} {820}},\
  \bibinfo {pages} {136484} (\bibinfo {year} {2021})},\ \Eprint
  {http://arxiv.org/abs/2012.10492} {arXiv:2012.10492 [hep-ph]} \BibitemShut
  {NoStop}%
\bibitem [{\citenamefont {Fuentes-Martin}\ \emph {et~al.}(2022)\citenamefont
  {Fuentes-Martin}, \citenamefont {Isidori}, \citenamefont {Lizana},
  \citenamefont {Selimovic},\ and\ \citenamefont
  {Stefanek}}]{Fuentes-Martin:2022xnb}%
  \BibitemOpen
  \bibfield  {author} {\bibinfo {author} {\bibfnamefont {J.}~\bibnamefont
  {Fuentes-Martin}}, \bibinfo {author} {\bibfnamefont {G.}~\bibnamefont
  {Isidori}}, \bibinfo {author} {\bibfnamefont {J.~M.}\ \bibnamefont {Lizana}},
  \bibinfo {author} {\bibfnamefont {N.}~\bibnamefont {Selimovic}}, \ and\
  \bibinfo {author} {\bibfnamefont {B.~A.}\ \bibnamefont {Stefanek}},\ }\href
  {\doibase 10.1016/j.physletb.2022.137382} {\bibfield  {journal} {\bibinfo
  {journal} {Phys. Lett. B}\ }\textbf {\bibinfo {volume} {834}},\ \bibinfo
  {pages} {137382} (\bibinfo {year} {2022})},\ \Eprint
  {http://arxiv.org/abs/2203.01952} {arXiv:2203.01952 [hep-ph]} \BibitemShut
  {NoStop}%
\bibitem [{\citenamefont {Assad}\ \emph {et~al.}(2018)\citenamefont {Assad},
  \citenamefont {Fornal},\ and\ \citenamefont {Grinstein}}]{Assad:2017iib}%
  \BibitemOpen
  \bibfield  {author} {\bibinfo {author} {\bibfnamefont {N.}~\bibnamefont
  {Assad}}, \bibinfo {author} {\bibfnamefont {B.}~\bibnamefont {Fornal}}, \
  and\ \bibinfo {author} {\bibfnamefont {B.}~\bibnamefont {Grinstein}},\ }\href
  {\doibase 10.1016/j.physletb.2017.12.042} {\bibfield  {journal} {\bibinfo
  {journal} {Phys. Lett. B}\ }\textbf {\bibinfo {volume} {777}},\ \bibinfo
  {pages} {324} (\bibinfo {year} {2018})},\ \Eprint
  {http://arxiv.org/abs/1708.06350} {arXiv:1708.06350 [hep-ph]} \BibitemShut
  {NoStop}%
\bibitem [{\citenamefont {Calibbi}\ \emph {et~al.}(2018)\citenamefont
  {Calibbi}, \citenamefont {Crivellin},\ and\ \citenamefont
  {Li}}]{Calibbi:2017qbu}%
  \BibitemOpen
  \bibfield  {author} {\bibinfo {author} {\bibfnamefont {L.}~\bibnamefont
  {Calibbi}}, \bibinfo {author} {\bibfnamefont {A.}~\bibnamefont {Crivellin}},
  \ and\ \bibinfo {author} {\bibfnamefont {T.}~\bibnamefont {Li}},\ }\href
  {\doibase 10.1103/PhysRevD.98.115002} {\bibfield  {journal} {\bibinfo
  {journal} {Phys. Rev. D}\ }\textbf {\bibinfo {volume} {98}},\ \bibinfo
  {pages} {115002} (\bibinfo {year} {2018})},\ \Eprint
  {http://arxiv.org/abs/1709.00692} {arXiv:1709.00692 [hep-ph]} \BibitemShut
  {NoStop}%
\bibitem [{\citenamefont {Barbieri}\ and\ \citenamefont
  {Tesi}(2018)}]{Barbieri:2017tuq}%
  \BibitemOpen
  \bibfield  {author} {\bibinfo {author} {\bibfnamefont {R.}~\bibnamefont
  {Barbieri}}\ and\ \bibinfo {author} {\bibfnamefont {A.}~\bibnamefont
  {Tesi}},\ }\href {\doibase 10.1140/epjc/s10052-018-5680-9} {\bibfield
  {journal} {\bibinfo  {journal} {Eur. Phys. J. C}\ }\textbf {\bibinfo {volume}
  {78}},\ \bibinfo {pages} {193} (\bibinfo {year} {2018})},\ \Eprint
  {http://arxiv.org/abs/1712.06844} {arXiv:1712.06844 [hep-ph]} \BibitemShut
  {NoStop}%
\bibitem [{\citenamefont {Blanke}\ and\ \citenamefont
  {Crivellin}(2018)}]{Blanke:2018sro}%
  \BibitemOpen
  \bibfield  {author} {\bibinfo {author} {\bibfnamefont {M.}~\bibnamefont
  {Blanke}}\ and\ \bibinfo {author} {\bibfnamefont {A.}~\bibnamefont
  {Crivellin}},\ }\href {\doibase 10.1103/PhysRevLett.121.011801} {\bibfield
  {journal} {\bibinfo  {journal} {Phys. Rev. Lett.}\ }\textbf {\bibinfo
  {volume} {121}},\ \bibinfo {pages} {011801} (\bibinfo {year} {2018})},\
  \Eprint {http://arxiv.org/abs/1801.07256} {arXiv:1801.07256 [hep-ph]}
  \BibitemShut {NoStop}%
\bibitem [{\citenamefont {Balaji}\ \emph {et~al.}(2019)\citenamefont {Balaji},
  \citenamefont {Foot},\ and\ \citenamefont {Schmidt}}]{Balaji:2018zna}%
  \BibitemOpen
  \bibfield  {author} {\bibinfo {author} {\bibfnamefont {S.}~\bibnamefont
  {Balaji}}, \bibinfo {author} {\bibfnamefont {R.}~\bibnamefont {Foot}}, \ and\
  \bibinfo {author} {\bibfnamefont {M.~A.}\ \bibnamefont {Schmidt}},\ }\href
  {\doibase 10.1103/PhysRevD.99.015029} {\bibfield  {journal} {\bibinfo
  {journal} {Phys. Rev. D}\ }\textbf {\bibinfo {volume} {99}},\ \bibinfo
  {pages} {015029} (\bibinfo {year} {2019})},\ \Eprint
  {http://arxiv.org/abs/1809.07562} {arXiv:1809.07562 [hep-ph]} \BibitemShut
  {NoStop}%
\bibitem [{\citenamefont {Dolan}\ \emph {et~al.}(2021)\citenamefont {Dolan},
  \citenamefont {Dutka},\ and\ \citenamefont {Volkas}}]{Dolan:2020doe}%
  \BibitemOpen
  \bibfield  {author} {\bibinfo {author} {\bibfnamefont {M.~J.}\ \bibnamefont
  {Dolan}}, \bibinfo {author} {\bibfnamefont {T.~P.}\ \bibnamefont {Dutka}}, \
  and\ \bibinfo {author} {\bibfnamefont {R.~R.}\ \bibnamefont {Volkas}},\
  }\href {\doibase 10.1007/JHEP05(2021)199} {\bibfield  {journal} {\bibinfo
  {journal} {JHEP}\ }\textbf {\bibinfo {volume} {05}},\ \bibinfo {pages} {199}
  (\bibinfo {year} {2021})},\ \Eprint {http://arxiv.org/abs/2012.05976}
  {arXiv:2012.05976 [hep-ph]} \BibitemShut {NoStop}%
\bibitem [{\citenamefont {King}(2021)}]{King:2021jeo}%
  \BibitemOpen
  \bibfield  {author} {\bibinfo {author} {\bibfnamefont {S.~F.}\ \bibnamefont
  {King}},\ }\href {\doibase 10.1007/JHEP11(2021)161} {\bibfield  {journal}
  {\bibinfo  {journal} {JHEP}\ }\textbf {\bibinfo {volume} {11}},\ \bibinfo
  {pages} {161} (\bibinfo {year} {2021})},\ \Eprint
  {http://arxiv.org/abs/2106.03876} {arXiv:2106.03876 [hep-ph]} \BibitemShut
  {NoStop}%
\bibitem [{\citenamefont {Fern\'andez~Navarro}\ and\ \citenamefont
  {King}(2022)}]{FernandezNavarro:2022gst}%
  \BibitemOpen
  \bibfield  {author} {\bibinfo {author} {\bibfnamefont {M.}~\bibnamefont
  {Fern\'andez~Navarro}}\ and\ \bibinfo {author} {\bibfnamefont {S.~F.}\
  \bibnamefont {King}},\ }\href@noop {} {\  (\bibinfo {year} {2022})},\ \Eprint
  {http://arxiv.org/abs/2209.00276} {arXiv:2209.00276 [hep-ph]} \BibitemShut
  {NoStop}%
\bibitem [{\citenamefont {Angelescu}\ \emph {et~al.}(2021)\citenamefont
  {Angelescu}, \citenamefont {Be\v{c}irevi\'c}, \citenamefont {Faroughy},
  \citenamefont {Jaffredo},\ and\ \citenamefont
  {Sumensari}}]{Angelescu:2021lln}%
  \BibitemOpen
  \bibfield  {author} {\bibinfo {author} {\bibfnamefont {A.}~\bibnamefont
  {Angelescu}}, \bibinfo {author} {\bibfnamefont {D.}~\bibnamefont
  {Be\v{c}irevi\'c}}, \bibinfo {author} {\bibfnamefont {D.~A.}\ \bibnamefont
  {Faroughy}}, \bibinfo {author} {\bibfnamefont {F.}~\bibnamefont {Jaffredo}},
  \ and\ \bibinfo {author} {\bibfnamefont {O.}~\bibnamefont {Sumensari}},\
  }\href {\doibase 10.1103/PhysRevD.104.055017} {\bibfield  {journal} {\bibinfo
   {journal} {Phys. Rev. D}\ }\textbf {\bibinfo {volume} {104}},\ \bibinfo
  {pages} {055017} (\bibinfo {year} {2021})},\ \Eprint
  {http://arxiv.org/abs/2103.12504} {arXiv:2103.12504 [hep-ph]} \BibitemShut
  {NoStop}%
\bibitem [{\citenamefont {Bhaskar}\ \emph {et~al.}(2021)\citenamefont
  {Bhaskar}, \citenamefont {Das}, \citenamefont {Mandal}, \citenamefont
  {Mitra},\ and\ \citenamefont {Neeraj}}]{Bhaskar:2021pml}%
  \BibitemOpen
  \bibfield  {author} {\bibinfo {author} {\bibfnamefont {A.}~\bibnamefont
  {Bhaskar}}, \bibinfo {author} {\bibfnamefont {D.}~\bibnamefont {Das}},
  \bibinfo {author} {\bibfnamefont {T.}~\bibnamefont {Mandal}}, \bibinfo
  {author} {\bibfnamefont {S.}~\bibnamefont {Mitra}}, \ and\ \bibinfo {author}
  {\bibfnamefont {C.}~\bibnamefont {Neeraj}},\ }\href {\doibase
  10.1103/PhysRevD.104.035016} {\bibfield  {journal} {\bibinfo  {journal}
  {Phys. Rev. D}\ }\textbf {\bibinfo {volume} {104}},\ \bibinfo {pages}
  {035016} (\bibinfo {year} {2021})},\ \Eprint
  {http://arxiv.org/abs/2101.12069} {arXiv:2101.12069 [hep-ph]} \BibitemShut
  {NoStop}%
\bibitem [{\citenamefont {Barbieri}\ \emph {et~al.}(2022)\citenamefont
  {Barbieri}, \citenamefont {Cornella},\ and\ \citenamefont
  {Isidori}}]{Barbieri:2022ikw}%
  \BibitemOpen
  \bibfield  {author} {\bibinfo {author} {\bibfnamefont {R.}~\bibnamefont
  {Barbieri}}, \bibinfo {author} {\bibfnamefont {C.}~\bibnamefont {Cornella}},
  \ and\ \bibinfo {author} {\bibfnamefont {G.}~\bibnamefont {Isidori}},\
  }\href@noop {} {\  (\bibinfo {year} {2022})},\ \Eprint
  {http://arxiv.org/abs/2207.14248} {arXiv:2207.14248 [hep-ph]} \BibitemShut
  {NoStop}%
\bibitem [{\citenamefont {Haisch}\ \emph {et~al.}(2022)\citenamefont {Haisch},
  \citenamefont {Schnell},\ and\ \citenamefont {Schulte}}]{Haisch:2022afh}%
  \BibitemOpen
  \bibfield  {author} {\bibinfo {author} {\bibfnamefont {U.}~\bibnamefont
  {Haisch}}, \bibinfo {author} {\bibfnamefont {L.}~\bibnamefont {Schnell}}, \
  and\ \bibinfo {author} {\bibfnamefont {S.}~\bibnamefont {Schulte}},\
  }\href@noop {} {\  (\bibinfo {year} {2022})},\ \Eprint
  {http://arxiv.org/abs/2209.12780} {arXiv:2209.12780 [hep-ph]} \BibitemShut
  {NoStop}%
\bibitem [{\citenamefont {Ciezarek}()}]{LHCbRd}%
  \BibitemOpen
  \bibfield  {author} {\bibinfo {author} {\bibfnamefont {G.~M.}\ \bibnamefont
  {Ciezarek}} (\bibinfo {collaboration} {LHCb}),\ }\href@noop {} {\bibinfo
  {journal} {CERN Seminar, https://indico.cern.ch/event/1187939/}\
  }\BibitemShut {NoStop}%
\bibitem [{\citenamefont {{The CMS
  Collaboration}}(2022{\natexlab{a}})}]{CMS:2022goy}%
  \BibitemOpen
\bibfield  {journal} {  }\bibfield  {author} {\bibinfo {author} {\bibnamefont
  {{The CMS Collaboration}}} (\bibinfo {collaboration} {CMS}),\ }\href@noop {}
  {\  (\bibinfo {year} {2022}{\natexlab{a}})},\ \Eprint
  {http://arxiv.org/abs/2208.02717} {arXiv:2208.02717 [hep-ex]} \BibitemShut
  {NoStop}%
\bibitem [{\citenamefont {{The CMS
  Collaboration}}(2022{\natexlab{b}})}]{CMS:2022zks}%
  \BibitemOpen
  \bibfield  {author} {\bibinfo {author} {\bibnamefont {{The CMS
  Collaboration}}} (\bibinfo {collaboration} {CMS}),\ }\href
  {https://cds.cern.ch/record/2815309} {\bibfield  {journal} {\bibinfo
  {journal} {CMS-PAS-EXO-19-016}\ } (\bibinfo {year}
  {2022}{\natexlab{b}})}\BibitemShut {NoStop}%
\bibitem [{\citenamefont {Faroughy}\ \emph {et~al.}(2017)\citenamefont
  {Faroughy}, \citenamefont {Greljo},\ and\ \citenamefont
  {Kamenik}}]{Faroughy:2016osc}%
  \BibitemOpen
  \bibfield  {author} {\bibinfo {author} {\bibfnamefont {D.~A.}\ \bibnamefont
  {Faroughy}}, \bibinfo {author} {\bibfnamefont {A.}~\bibnamefont {Greljo}}, \
  and\ \bibinfo {author} {\bibfnamefont {J.~F.}\ \bibnamefont {Kamenik}},\
  }\href {\doibase 10.1016/j.physletb.2016.11.011} {\bibfield  {journal}
  {\bibinfo  {journal} {Phys. Lett. B}\ }\textbf {\bibinfo {volume} {764}},\
  \bibinfo {pages} {126} (\bibinfo {year} {2017})},\ \Eprint
  {http://arxiv.org/abs/1609.07138} {arXiv:1609.07138 [hep-ph]} \BibitemShut
  {NoStop}%
\bibitem [{\citenamefont {Aad}\ \emph {et~al.}(2020)\citenamefont {Aad} \emph
  {et~al.}}]{ATLAS:2020zms}%
  \BibitemOpen
  \bibfield  {author} {\bibinfo {author} {\bibfnamefont {G.}~\bibnamefont
  {Aad}} \emph {et~al.} (\bibinfo {collaboration} {ATLAS}),\ }\href {\doibase
  10.1103/PhysRevLett.125.051801} {\bibfield  {journal} {\bibinfo  {journal}
  {Phys. Rev. Lett.}\ }\textbf {\bibinfo {volume} {125}},\ \bibinfo {pages}
  {051801} (\bibinfo {year} {2020})},\ \Eprint
  {http://arxiv.org/abs/2002.12223} {arXiv:2002.12223 [hep-ex]} \BibitemShut
  {NoStop}%
\bibitem [{\citenamefont {Fuentes-Mart\'\i{}n}\ \emph
  {et~al.}(2020{\natexlab{c}})\citenamefont {Fuentes-Mart\'\i{}n},
  \citenamefont {Isidori}, \citenamefont {Pag\`es},\ and\ \citenamefont
  {Yamamoto}}]{Fuentes-Martin:2019mun}%
  \BibitemOpen
  \bibfield  {author} {\bibinfo {author} {\bibfnamefont {J.}~\bibnamefont
  {Fuentes-Mart\'\i{}n}}, \bibinfo {author} {\bibfnamefont {G.}~\bibnamefont
  {Isidori}}, \bibinfo {author} {\bibfnamefont {J.}~\bibnamefont {Pag\`es}}, \
  and\ \bibinfo {author} {\bibfnamefont {K.}~\bibnamefont {Yamamoto}},\ }\href
  {\doibase 10.1016/j.physletb.2019.135080} {\bibfield  {journal} {\bibinfo
  {journal} {Phys. Lett. B}\ }\textbf {\bibinfo {volume} {800}},\ \bibinfo
  {pages} {135080} (\bibinfo {year} {2020}{\natexlab{c}})},\ \Eprint
  {http://arxiv.org/abs/1909.02519} {arXiv:1909.02519 [hep-ph]} \BibitemShut
  {NoStop}%
\bibitem [{\citenamefont {Barbieri}\ \emph {et~al.}(2011)\citenamefont
  {Barbieri}, \citenamefont {Isidori}, \citenamefont {Jones-Perez},
  \citenamefont {Lodone},\ and\ \citenamefont {Straub}}]{Barbieri:2011ci}%
  \BibitemOpen
  \bibfield  {author} {\bibinfo {author} {\bibfnamefont {R.}~\bibnamefont
  {Barbieri}}, \bibinfo {author} {\bibfnamefont {G.}~\bibnamefont {Isidori}},
  \bibinfo {author} {\bibfnamefont {J.}~\bibnamefont {Jones-Perez}}, \bibinfo
  {author} {\bibfnamefont {P.}~\bibnamefont {Lodone}}, \ and\ \bibinfo {author}
  {\bibfnamefont {D.~M.}\ \bibnamefont {Straub}},\ }\href {\doibase
  10.1140/epjc/s10052-011-1725-z} {\bibfield  {journal} {\bibinfo  {journal}
  {Eur. Phys. J. C}\ }\textbf {\bibinfo {volume} {71}},\ \bibinfo {pages}
  {1725} (\bibinfo {year} {2011})},\ \Eprint {http://arxiv.org/abs/1105.2296}
  {arXiv:1105.2296 [hep-ph]} \BibitemShut {NoStop}%
\bibitem [{\citenamefont {Barbieri}\ \emph {et~al.}(2012)\citenamefont
  {Barbieri}, \citenamefont {Buttazzo}, \citenamefont {Sala},\ and\
  \citenamefont {Straub}}]{Barbieri:2012uh}%
  \BibitemOpen
  \bibfield  {author} {\bibinfo {author} {\bibfnamefont {R.}~\bibnamefont
  {Barbieri}}, \bibinfo {author} {\bibfnamefont {D.}~\bibnamefont {Buttazzo}},
  \bibinfo {author} {\bibfnamefont {F.}~\bibnamefont {Sala}}, \ and\ \bibinfo
  {author} {\bibfnamefont {D.~M.}\ \bibnamefont {Straub}},\ }\href {\doibase
  10.1007/JHEP07(2012)181} {\bibfield  {journal} {\bibinfo  {journal} {JHEP}\
  }\textbf {\bibinfo {volume} {07}},\ \bibinfo {pages} {181} (\bibinfo {year}
  {2012})},\ \Eprint {http://arxiv.org/abs/1203.4218} {arXiv:1203.4218
  [hep-ph]} \BibitemShut {NoStop}%
\bibitem [{\citenamefont {Crosas}\ \emph {et~al.}(2022)\citenamefont {Crosas},
  \citenamefont {Isidori}, \citenamefont {Lizana}, \citenamefont {Selimovic},\
  and\ \citenamefont {Stefanek}}]{Crosas:2022quq}%
  \BibitemOpen
  \bibfield  {author} {\bibinfo {author} {\bibfnamefont {O.~L.}\ \bibnamefont
  {Crosas}}, \bibinfo {author} {\bibfnamefont {G.}~\bibnamefont {Isidori}},
  \bibinfo {author} {\bibfnamefont {J.~M.}\ \bibnamefont {Lizana}}, \bibinfo
  {author} {\bibfnamefont {N.}~\bibnamefont {Selimovic}}, \ and\ \bibinfo
  {author} {\bibfnamefont {B.~A.}\ \bibnamefont {Stefanek}},\ }\href@noop {} {\
   (\bibinfo {year} {2022})},\ \Eprint {http://arxiv.org/abs/2207.00018}
  {arXiv:2207.00018 [hep-ph]} \BibitemShut {NoStop}%
\bibitem [{\citenamefont {Baker}\ \emph {et~al.}(2019)\citenamefont {Baker},
  \citenamefont {Fuentes-Mart\'\i{}n}, \citenamefont {Isidori},\ and\
  \citenamefont {K\"onig}}]{Baker:2019sli}%
  \BibitemOpen
  \bibfield  {author} {\bibinfo {author} {\bibfnamefont {M.~J.}\ \bibnamefont
  {Baker}}, \bibinfo {author} {\bibfnamefont {J.}~\bibnamefont
  {Fuentes-Mart\'\i{}n}}, \bibinfo {author} {\bibfnamefont {G.}~\bibnamefont
  {Isidori}}, \ and\ \bibinfo {author} {\bibfnamefont {M.}~\bibnamefont
  {K\"onig}},\ }\href {\doibase 10.1140/epjc/s10052-019-6853-x} {\bibfield
  {journal} {\bibinfo  {journal} {Eur. Phys. J. C}\ }\textbf {\bibinfo {volume}
  {79}},\ \bibinfo {pages} {334} (\bibinfo {year} {2019})},\ \Eprint
  {http://arxiv.org/abs/1901.10480} {arXiv:1901.10480 [hep-ph]} \BibitemShut
  {NoStop}%
\bibitem [{\citenamefont {Alonso}\ \emph {et~al.}(2017)\citenamefont {Alonso},
  \citenamefont {Grinstein},\ and\ \citenamefont
  {Martin~Camalich}}]{Alonso:2016oyd}%
  \BibitemOpen
  \bibfield  {author} {\bibinfo {author} {\bibfnamefont {R.}~\bibnamefont
  {Alonso}}, \bibinfo {author} {\bibfnamefont {B.}~\bibnamefont {Grinstein}}, \
  and\ \bibinfo {author} {\bibfnamefont {J.}~\bibnamefont {Martin~Camalich}},\
  }\href {\doibase 10.1103/PhysRevLett.118.081802} {\bibfield  {journal}
  {\bibinfo  {journal} {Phys. Rev. Lett.}\ }\textbf {\bibinfo {volume} {118}},\
  \bibinfo {pages} {081802} (\bibinfo {year} {2017})},\ \Eprint
  {http://arxiv.org/abs/1611.06676} {arXiv:1611.06676 [hep-ph]} \BibitemShut
  {NoStop}%
\bibitem [{\citenamefont {Aaij}\ \emph
  {et~al.}(2023{\natexlab{a}})\citenamefont {Aaij} \emph
  {et~al.}}]{LHCb:2023zxo}%
  \BibitemOpen
  \bibfield  {author} {\bibinfo {author} {\bibfnamefont {R.}~\bibnamefont
  {Aaij}} \emph {et~al.} (\bibinfo {collaboration} {LHCb}),\ }\href@noop {} {\
  (\bibinfo {year} {2023}{\natexlab{a}})},\ \Eprint
  {http://arxiv.org/abs/2302.02886} {arXiv:2302.02886 [hep-ex]} \BibitemShut
  {NoStop}%
\bibitem [{\citenamefont {Aaij}\ \emph
  {et~al.}(2023{\natexlab{b}})\citenamefont {Aaij} \emph
  {et~al.}}]{CERNRDstar}%
  \BibitemOpen
  \bibfield  {author} {\bibinfo {author} {\bibfnamefont {R.}~\bibnamefont
  {Aaij}} \emph {et~al.} (\bibinfo {collaboration} {LHCb}),\ }\href@noop {}
  {\bibfield  {journal} {\bibinfo  {journal}
  {https://indico.cern.ch/event/1231797}\ } (\bibinfo {year}
  {2023}{\natexlab{b}})}\BibitemShut {NoStop}%
\bibitem [{\citenamefont {Amhis}\ \emph {et~al.}(2021)\citenamefont {Amhis}
  \emph {et~al.}}]{HFLAV:2019otj}%
  \BibitemOpen
  \bibfield  {author} {\bibinfo {author} {\bibfnamefont {Y.~S.}\ \bibnamefont
  {Amhis}} \emph {et~al.} (\bibinfo {collaboration} {HFLAV
  [https://hflav.web.cern.ch]}),\ }\href {\doibase
  10.1140/epjc/s10052-020-8156-7} {\bibfield  {journal} {\bibinfo  {journal}
  {Eur. Phys. J. C}\ }\textbf {\bibinfo {volume} {81}},\ \bibinfo {pages} {226}
  (\bibinfo {year} {2021})},\ \Eprint {http://arxiv.org/abs/1909.12524}
  {arXiv:1909.12524 [hep-ex]} \BibitemShut {NoStop}%
\bibitem [{\citenamefont {Bailey}\ \emph {et~al.}(2015)\citenamefont {Bailey}
  \emph {et~al.}}]{MILC:2015uhg}%
  \BibitemOpen
  \bibfield  {author} {\bibinfo {author} {\bibfnamefont {J.~A.}\ \bibnamefont
  {Bailey}} \emph {et~al.} (\bibinfo {collaboration} {MILC}),\ }\href {\doibase
  10.1103/PhysRevD.92.034506} {\bibfield  {journal} {\bibinfo  {journal} {Phys.
  Rev. D}\ }\textbf {\bibinfo {volume} {92}},\ \bibinfo {pages} {034506}
  (\bibinfo {year} {2015})},\ \Eprint {http://arxiv.org/abs/1503.07237}
  {arXiv:1503.07237 [hep-lat]} \BibitemShut {NoStop}%
\bibitem [{\citenamefont {Na}\ \emph {et~al.}(2015)\citenamefont {Na},
  \citenamefont {Bouchard}, \citenamefont {Lepage}, \citenamefont {Monahan},\
  and\ \citenamefont {Shigemitsu}}]{Na:2015kha}%
  \BibitemOpen
  \bibfield  {author} {\bibinfo {author} {\bibfnamefont {H.}~\bibnamefont
  {Na}}, \bibinfo {author} {\bibfnamefont {C.~M.}\ \bibnamefont {Bouchard}},
  \bibinfo {author} {\bibfnamefont {G.~P.}\ \bibnamefont {Lepage}}, \bibinfo
  {author} {\bibfnamefont {C.}~\bibnamefont {Monahan}}, \ and\ \bibinfo
  {author} {\bibfnamefont {J.}~\bibnamefont {Shigemitsu}} (\bibinfo
  {collaboration} {HPQCD}),\ }\href {\doibase 10.1103/PhysRevD.93.119906}
  {\bibfield  {journal} {\bibinfo  {journal} {Phys. Rev. D}\ }\textbf {\bibinfo
  {volume} {92}},\ \bibinfo {pages} {054510} (\bibinfo {year} {2015})},\
  \bibinfo {note} {[Erratum: Phys.Rev.D 93, 119906 (2016)]},\ \Eprint
  {http://arxiv.org/abs/1505.03925} {arXiv:1505.03925 [hep-lat]} \BibitemShut
  {NoStop}%
\bibitem [{\citenamefont {Bernlochner}\ \emph {et~al.}(2017)\citenamefont
  {Bernlochner}, \citenamefont {Ligeti}, \citenamefont {Papucci},\ and\
  \citenamefont {Robinson}}]{Bernlochner:2017jka}%
  \BibitemOpen
  \bibfield  {author} {\bibinfo {author} {\bibfnamefont {F.~U.}\ \bibnamefont
  {Bernlochner}}, \bibinfo {author} {\bibfnamefont {Z.}~\bibnamefont {Ligeti}},
  \bibinfo {author} {\bibfnamefont {M.}~\bibnamefont {Papucci}}, \ and\
  \bibinfo {author} {\bibfnamefont {D.~J.}\ \bibnamefont {Robinson}},\ }\href
  {\doibase 10.1103/PhysRevD.95.115008} {\bibfield  {journal} {\bibinfo
  {journal} {Phys. Rev. D}\ }\textbf {\bibinfo {volume} {95}},\ \bibinfo
  {pages} {115008} (\bibinfo {year} {2017})},\ \bibinfo {note} {[Erratum:
  Phys.Rev.D 97, 059902 (2018)]},\ \Eprint {http://arxiv.org/abs/1703.05330}
  {arXiv:1703.05330 [hep-ph]} \BibitemShut {NoStop}%
\bibitem [{\citenamefont {Gambino}\ \emph {et~al.}(2019)\citenamefont
  {Gambino}, \citenamefont {Jung},\ and\ \citenamefont
  {Schacht}}]{Gambino:2019sif}%
  \BibitemOpen
  \bibfield  {author} {\bibinfo {author} {\bibfnamefont {P.}~\bibnamefont
  {Gambino}}, \bibinfo {author} {\bibfnamefont {M.}~\bibnamefont {Jung}}, \
  and\ \bibinfo {author} {\bibfnamefont {S.}~\bibnamefont {Schacht}},\ }\href
  {\doibase 10.1016/j.physletb.2019.06.039} {\bibfield  {journal} {\bibinfo
  {journal} {Phys. Lett. B}\ }\textbf {\bibinfo {volume} {795}},\ \bibinfo
  {pages} {386} (\bibinfo {year} {2019})},\ \Eprint
  {http://arxiv.org/abs/1905.08209} {arXiv:1905.08209 [hep-ph]} \BibitemShut
  {NoStop}%
\bibitem [{\citenamefont {Bordone}\ \emph {et~al.}(2020)\citenamefont
  {Bordone}, \citenamefont {Jung},\ and\ \citenamefont {van
  Dyk}}]{Bordone:2019vic}%
  \BibitemOpen
  \bibfield  {author} {\bibinfo {author} {\bibfnamefont {M.}~\bibnamefont
  {Bordone}}, \bibinfo {author} {\bibfnamefont {M.}~\bibnamefont {Jung}}, \
  and\ \bibinfo {author} {\bibfnamefont {D.}~\bibnamefont {van Dyk}},\ }\href
  {\doibase 10.1140/epjc/s10052-020-7616-4} {\bibfield  {journal} {\bibinfo
  {journal} {Eur. Phys. J. C}\ }\textbf {\bibinfo {volume} {80}},\ \bibinfo
  {pages} {74} (\bibinfo {year} {2020})},\ \Eprint
  {http://arxiv.org/abs/1908.09398} {arXiv:1908.09398 [hep-ph]} \BibitemShut
  {NoStop}%
\bibitem [{\citenamefont {Martinelli}\ \emph {et~al.}(2022)\citenamefont
  {Martinelli}, \citenamefont {Simula},\ and\ \citenamefont
  {Vittorio}}]{Martinelli:2021onb}%
  \BibitemOpen
  \bibfield  {author} {\bibinfo {author} {\bibfnamefont {G.}~\bibnamefont
  {Martinelli}}, \bibinfo {author} {\bibfnamefont {S.}~\bibnamefont {Simula}},
  \ and\ \bibinfo {author} {\bibfnamefont {L.}~\bibnamefont {Vittorio}},\
  }\href {\doibase 10.1103/PhysRevD.105.034503} {\bibfield  {journal} {\bibinfo
   {journal} {Phys. Rev. D}\ }\textbf {\bibinfo {volume} {105}},\ \bibinfo
  {pages} {034503} (\bibinfo {year} {2022})},\ \Eprint
  {http://arxiv.org/abs/2105.08674} {arXiv:2105.08674 [hep-ph]} \BibitemShut
  {NoStop}%
\bibitem [{\citenamefont {Be\v{c}irevi\'c}\ and\ \citenamefont
  {Jaffredo}(2022)}]{Becirevic:2022bev}%
  \BibitemOpen
  \bibfield  {author} {\bibinfo {author} {\bibfnamefont {D.}~\bibnamefont
  {Be\v{c}irevi\'c}}\ and\ \bibinfo {author} {\bibfnamefont {F.}~\bibnamefont
  {Jaffredo}},\ }\href@noop {} {\  (\bibinfo {year} {2022})},\ \Eprint
  {http://arxiv.org/abs/2209.13409} {arXiv:2209.13409 [hep-ph]} \BibitemShut
  {NoStop}%
\bibitem [{\citenamefont {Aaij}\ \emph {et~al.}(2022)\citenamefont {Aaij} \emph
  {et~al.}}]{LHCb:2022piu}%
  \BibitemOpen
  \bibfield  {author} {\bibinfo {author} {\bibfnamefont {R.}~\bibnamefont
  {Aaij}} \emph {et~al.} (\bibinfo {collaboration} {LHCb}),\ }\href {\doibase
  10.1103/PhysRevLett.128.191803} {\bibfield  {journal} {\bibinfo  {journal}
  {Phys. Rev. Lett.}\ }\textbf {\bibinfo {volume} {128}},\ \bibinfo {pages}
  {191803} (\bibinfo {year} {2022})},\ \Eprint
  {http://arxiv.org/abs/2201.03497} {arXiv:2201.03497 [hep-ex]} \BibitemShut
  {NoStop}%
\bibitem [{\citenamefont {Workman}\ \emph {et~al.}(2022)\citenamefont {Workman}
  \emph {et~al.}}]{ParticleDataGroup:2022pth}%
  \BibitemOpen
  \bibfield  {author} {\bibinfo {author} {\bibfnamefont {R.~L.}\ \bibnamefont
  {Workman}} \emph {et~al.} (\bibinfo {collaboration} {Particle Data Group}),\
  }\href {\doibase 10.1093/ptep/ptac097} {\bibfield  {journal} {\bibinfo
  {journal} {PTEP}\ }\textbf {\bibinfo {volume} {2022}},\ \bibinfo {pages}
  {083C01} (\bibinfo {year} {2022})}\BibitemShut {NoStop}%
\bibitem [{\citenamefont {Bona}\ \emph {et~al.}(2022)\citenamefont {Bona} \emph
  {et~al.}}]{Bona:2022zhn}%
  \BibitemOpen
  \bibfield  {author} {\bibinfo {author} {\bibfnamefont {M.}~\bibnamefont
  {Bona}} \emph {et~al.} (\bibinfo {collaboration} {UTfit
  [http://www.utfit.org/UTfit/]}),\ }\href {\doibase 10.22323/1.398.0500}
  {\bibfield  {journal} {\bibinfo  {journal} {PoS}\ }\textbf {\bibinfo {volume}
  {EPS-HEP2021}},\ \bibinfo {pages} {500} (\bibinfo {year} {2022})}\BibitemShut
  {NoStop}%
\bibitem [{\citenamefont {Aebischer}\ \emph {et~al.}(2017)\citenamefont
  {Aebischer}, \citenamefont {Fael}, \citenamefont {Greub},\ and\ \citenamefont
  {Virto}}]{Aebischer:2017gaw}%
  \BibitemOpen
  \bibfield  {author} {\bibinfo {author} {\bibfnamefont {J.}~\bibnamefont
  {Aebischer}}, \bibinfo {author} {\bibfnamefont {M.}~\bibnamefont {Fael}},
  \bibinfo {author} {\bibfnamefont {C.}~\bibnamefont {Greub}}, \ and\ \bibinfo
  {author} {\bibfnamefont {J.}~\bibnamefont {Virto}},\ }\href {\doibase
  10.1007/JHEP09(2017)158} {\bibfield  {journal} {\bibinfo  {journal} {JHEP}\
  }\textbf {\bibinfo {volume} {09}},\ \bibinfo {pages} {158} (\bibinfo {year}
  {2017})},\ \Eprint {http://arxiv.org/abs/1704.06639} {arXiv:1704.06639
  [hep-ph]} \BibitemShut {NoStop}%
\bibitem [{\citenamefont {Crivellin}\ \emph {et~al.}(2019)\citenamefont
  {Crivellin}, \citenamefont {Greub}, \citenamefont {M\"uller},\ and\
  \citenamefont {Saturnino}}]{Crivellin:2018yvo}%
  \BibitemOpen
  \bibfield  {author} {\bibinfo {author} {\bibfnamefont {A.}~\bibnamefont
  {Crivellin}}, \bibinfo {author} {\bibfnamefont {C.}~\bibnamefont {Greub}},
  \bibinfo {author} {\bibfnamefont {D.}~\bibnamefont {M\"uller}}, \ and\
  \bibinfo {author} {\bibfnamefont {F.}~\bibnamefont {Saturnino}},\ }\href
  {\doibase 10.1103/PhysRevLett.122.011805} {\bibfield  {journal} {\bibinfo
  {journal} {Phys. Rev. Lett.}\ }\textbf {\bibinfo {volume} {122}},\ \bibinfo
  {pages} {011805} (\bibinfo {year} {2019})},\ \Eprint
  {http://arxiv.org/abs/1807.02068} {arXiv:1807.02068 [hep-ph]} \BibitemShut
  {NoStop}%
\bibitem [{\citenamefont {London}\ and\ \citenamefont
  {Matias}(2022)}]{London:2021lfn}%
  \BibitemOpen
  \bibfield  {author} {\bibinfo {author} {\bibfnamefont {D.}~\bibnamefont
  {London}}\ and\ \bibinfo {author} {\bibfnamefont {J.}~\bibnamefont
  {Matias}},\ }\href {\doibase 10.1146/annurev-nucl-102020-090209} {\bibfield
  {journal} {\bibinfo  {journal} {Ann. Rev. Nucl. Part. Sci.}\ }\textbf
  {\bibinfo {volume} {72}},\ \bibinfo {pages} {37} (\bibinfo {year} {2022})},\
  \Eprint {http://arxiv.org/abs/2110.13270} {arXiv:2110.13270 [hep-ph]}
  \BibitemShut {NoStop}%
\bibitem [{\citenamefont {Altmannshofer}\ and\ \citenamefont
  {Stangl}(2021)}]{Altmannshofer:2021qrr}%
  \BibitemOpen
  \bibfield  {author} {\bibinfo {author} {\bibfnamefont {W.}~\bibnamefont
  {Altmannshofer}}\ and\ \bibinfo {author} {\bibfnamefont {P.}~\bibnamefont
  {Stangl}},\ }\href {\doibase 10.1140/epjc/s10052-021-09725-1} {\bibfield
  {journal} {\bibinfo  {journal} {Eur. Phys. J. C}\ }\textbf {\bibinfo {volume}
  {81}},\ \bibinfo {pages} {952} (\bibinfo {year} {2021})},\ \Eprint
  {http://arxiv.org/abs/2103.13370} {arXiv:2103.13370 [hep-ph]} \BibitemShut
  {NoStop}%
\bibitem [{\citenamefont {Cornella}\ \emph {et~al.}(2020)\citenamefont
  {Cornella}, \citenamefont {Isidori}, \citenamefont {K\"onig}, \citenamefont
  {Liechti}, \citenamefont {Owen},\ and\ \citenamefont
  {Serra}}]{Cornella:2020aoq}%
  \BibitemOpen
  \bibfield  {author} {\bibinfo {author} {\bibfnamefont {C.}~\bibnamefont
  {Cornella}}, \bibinfo {author} {\bibfnamefont {G.}~\bibnamefont {Isidori}},
  \bibinfo {author} {\bibfnamefont {M.}~\bibnamefont {K\"onig}}, \bibinfo
  {author} {\bibfnamefont {S.}~\bibnamefont {Liechti}}, \bibinfo {author}
  {\bibfnamefont {P.}~\bibnamefont {Owen}}, \ and\ \bibinfo {author}
  {\bibfnamefont {N.}~\bibnamefont {Serra}},\ }\href {\doibase
  10.1140/epjc/s10052-020-08674-5} {\bibfield  {journal} {\bibinfo  {journal}
  {Eur. Phys. J. C}\ }\textbf {\bibinfo {volume} {80}},\ \bibinfo {pages}
  {1095} (\bibinfo {year} {2020})},\ \Eprint {http://arxiv.org/abs/2001.04470}
  {arXiv:2001.04470 [hep-ph]} \BibitemShut {NoStop}%
\bibitem [{\citenamefont {Khodjamirian}\ \emph {et~al.}(2013)\citenamefont
  {Khodjamirian}, \citenamefont {Mannel},\ and\ \citenamefont
  {Wang}}]{Khodjamirian:2012rm}%
  \BibitemOpen
  \bibfield  {author} {\bibinfo {author} {\bibfnamefont {A.}~\bibnamefont
  {Khodjamirian}}, \bibinfo {author} {\bibfnamefont {T.}~\bibnamefont
  {Mannel}}, \ and\ \bibinfo {author} {\bibfnamefont {Y.~M.}\ \bibnamefont
  {Wang}},\ }\href {\doibase 10.1007/JHEP02(2013)010} {\bibfield  {journal}
  {\bibinfo  {journal} {JHEP}\ }\textbf {\bibinfo {volume} {02}},\ \bibinfo
  {pages} {010} (\bibinfo {year} {2013})},\ \Eprint
  {http://arxiv.org/abs/1211.0234} {arXiv:1211.0234 [hep-ph]} \BibitemShut
  {NoStop}%
\bibitem [{\citenamefont {Allwicher}\ \emph
  {et~al.}(2022{\natexlab{a}})\citenamefont {Allwicher}, \citenamefont
  {Faroughy}, \citenamefont {Jaffredo}, \citenamefont {Sumensari},\ and\
  \citenamefont {Wilsch}}]{Allwicher:2022gkm}%
  \BibitemOpen
  \bibfield  {author} {\bibinfo {author} {\bibfnamefont {L.}~\bibnamefont
  {Allwicher}}, \bibinfo {author} {\bibfnamefont {D.~A.}\ \bibnamefont
  {Faroughy}}, \bibinfo {author} {\bibfnamefont {F.}~\bibnamefont {Jaffredo}},
  \bibinfo {author} {\bibfnamefont {O.}~\bibnamefont {Sumensari}}, \ and\
  \bibinfo {author} {\bibfnamefont {F.}~\bibnamefont {Wilsch}},\ }\href@noop {}
  {\  (\bibinfo {year} {2022}{\natexlab{a}})},\ \Eprint
  {http://arxiv.org/abs/2207.10714} {arXiv:2207.10714 [hep-ph]} \BibitemShut
  {NoStop}%
\bibitem [{\citenamefont {Diaz}\ \emph {et~al.}(2017)\citenamefont {Diaz},
  \citenamefont {Schmaltz},\ and\ \citenamefont {Zhong}}]{Diaz:2017lit}%
  \BibitemOpen
  \bibfield  {author} {\bibinfo {author} {\bibfnamefont {B.}~\bibnamefont
  {Diaz}}, \bibinfo {author} {\bibfnamefont {M.}~\bibnamefont {Schmaltz}}, \
  and\ \bibinfo {author} {\bibfnamefont {Y.-M.}\ \bibnamefont {Zhong}},\ }\href
  {\doibase 10.1007/JHEP10(2017)097} {\bibfield  {journal} {\bibinfo  {journal}
  {JHEP}\ }\textbf {\bibinfo {volume} {10}},\ \bibinfo {pages} {097} (\bibinfo
  {year} {2017})},\ \Eprint {http://arxiv.org/abs/1706.05033} {arXiv:1706.05033
  [hep-ph]} \BibitemShut {NoStop}%
\bibitem [{\citenamefont {Blumlein}\ \emph {et~al.}(1997)\citenamefont
  {Blumlein}, \citenamefont {Boos},\ and\ \citenamefont
  {Kryukov}}]{Blumlein:1996qp}%
  \BibitemOpen
  \bibfield  {author} {\bibinfo {author} {\bibfnamefont {J.}~\bibnamefont
  {Blumlein}}, \bibinfo {author} {\bibfnamefont {E.}~\bibnamefont {Boos}}, \
  and\ \bibinfo {author} {\bibfnamefont {A.}~\bibnamefont {Kryukov}},\ }\href
  {\doibase 10.1007/s002880050538} {\bibfield  {journal} {\bibinfo  {journal}
  {Z. Phys. C}\ }\textbf {\bibinfo {volume} {76}},\ \bibinfo {pages} {137}
  (\bibinfo {year} {1997})},\ \Eprint {http://arxiv.org/abs/hep-ph/9610408}
  {arXiv:hep-ph/9610408} \BibitemShut {NoStop}%
\bibitem [{\citenamefont {Dor\v{s}ner}\ and\ \citenamefont
  {Greljo}(2018)}]{Dorsner:2018ynv}%
  \BibitemOpen
  \bibfield  {author} {\bibinfo {author} {\bibfnamefont {I.}~\bibnamefont
  {Dor\v{s}ner}}\ and\ \bibinfo {author} {\bibfnamefont {A.}~\bibnamefont
  {Greljo}},\ }\href {\doibase 10.1007/JHEP05(2018)126} {\bibfield  {journal}
  {\bibinfo  {journal} {JHEP}\ }\textbf {\bibinfo {volume} {05}},\ \bibinfo
  {pages} {126} (\bibinfo {year} {2018})},\ \Eprint
  {http://arxiv.org/abs/1801.07641} {arXiv:1801.07641 [hep-ph]} \BibitemShut
  {NoStop}%
\bibitem [{\citenamefont {Aad}\ \emph {et~al.}(2021)\citenamefont {Aad} \emph
  {et~al.}}]{ATLAS:2021jyv}%
  \BibitemOpen
  \bibfield  {author} {\bibinfo {author} {\bibfnamefont {G.}~\bibnamefont
  {Aad}} \emph {et~al.} (\bibinfo {collaboration} {ATLAS}),\ }\href {\doibase
  10.1103/PhysRevD.104.112005} {\bibfield  {journal} {\bibinfo  {journal}
  {Phys. Rev. D}\ }\textbf {\bibinfo {volume} {104}},\ \bibinfo {pages}
  {112005} (\bibinfo {year} {2021})},\ \Eprint
  {http://arxiv.org/abs/2108.07665} {arXiv:2108.07665 [hep-ex]} \BibitemShut
  {NoStop}%
\bibitem [{\citenamefont {Sirunyan}\ \emph {et~al.}(2021)\citenamefont
  {Sirunyan} \emph {et~al.}}]{CMS:2020wzx}%
  \BibitemOpen
  \bibfield  {author} {\bibinfo {author} {\bibfnamefont {A.~M.}\ \bibnamefont
  {Sirunyan}} \emph {et~al.} (\bibinfo {collaboration} {CMS}),\ }\href
  {\doibase 10.1016/j.physletb.2021.136446} {\bibfield  {journal} {\bibinfo
  {journal} {Phys. Lett. B}\ }\textbf {\bibinfo {volume} {819}},\ \bibinfo
  {pages} {136446} (\bibinfo {year} {2021})},\ \Eprint
  {http://arxiv.org/abs/2012.04178} {arXiv:2012.04178 [hep-ex]} \BibitemShut
  {NoStop}%
\bibitem [{\citenamefont {Hammett}\ and\ \citenamefont
  {Ross}(2015)}]{Hammett:2015sea}%
  \BibitemOpen
  \bibfield  {author} {\bibinfo {author} {\bibfnamefont {J.~B.}\ \bibnamefont
  {Hammett}}\ and\ \bibinfo {author} {\bibfnamefont {D.~A.}\ \bibnamefont
  {Ross}},\ }\href {\doibase 10.1007/JHEP07(2015)148} {\bibfield  {journal}
  {\bibinfo  {journal} {JHEP}\ }\textbf {\bibinfo {volume} {07}},\ \bibinfo
  {pages} {148} (\bibinfo {year} {2015})},\ \Eprint
  {http://arxiv.org/abs/1501.06719} {arXiv:1501.06719 [hep-ph]} \BibitemShut
  {NoStop}%
\bibitem [{\citenamefont {Mandal}\ \emph {et~al.}(2015)\citenamefont {Mandal},
  \citenamefont {Mitra},\ and\ \citenamefont {Seth}}]{Mandal:2015vfa}%
  \BibitemOpen
  \bibfield  {author} {\bibinfo {author} {\bibfnamefont {T.}~\bibnamefont
  {Mandal}}, \bibinfo {author} {\bibfnamefont {S.}~\bibnamefont {Mitra}}, \
  and\ \bibinfo {author} {\bibfnamefont {S.}~\bibnamefont {Seth}},\ }\href
  {\doibase 10.1007/JHEP07(2015)028} {\bibfield  {journal} {\bibinfo  {journal}
  {JHEP}\ }\textbf {\bibinfo {volume} {07}},\ \bibinfo {pages} {028} (\bibinfo
  {year} {2015})},\ \Eprint {http://arxiv.org/abs/1503.04689} {arXiv:1503.04689
  [hep-ph]} \BibitemShut {NoStop}%
\bibitem [{\citenamefont {Alves}\ \emph {et~al.}(2003)\citenamefont {Alves},
  \citenamefont {Eboli},\ and\ \citenamefont {Plehn}}]{Alves:2002tj}%
  \BibitemOpen
  \bibfield  {author} {\bibinfo {author} {\bibfnamefont {A.}~\bibnamefont
  {Alves}}, \bibinfo {author} {\bibfnamefont {O.}~\bibnamefont {Eboli}}, \ and\
  \bibinfo {author} {\bibfnamefont {T.}~\bibnamefont {Plehn}},\ }\href
  {\doibase 10.1016/S0370-2693(03)00266-1} {\bibfield  {journal} {\bibinfo
  {journal} {Phys. Lett. B}\ }\textbf {\bibinfo {volume} {558}},\ \bibinfo
  {pages} {165} (\bibinfo {year} {2003})},\ \Eprint
  {http://arxiv.org/abs/hep-ph/0211441} {arXiv:hep-ph/0211441} \BibitemShut
  {NoStop}%
\bibitem [{\citenamefont {Haisch}\ and\ \citenamefont
  {Polesello}(2021)}]{Haisch:2020xjd}%
  \BibitemOpen
  \bibfield  {author} {\bibinfo {author} {\bibfnamefont {U.}~\bibnamefont
  {Haisch}}\ and\ \bibinfo {author} {\bibfnamefont {G.}~\bibnamefont
  {Polesello}},\ }\href {\doibase 10.1007/JHEP05(2021)057} {\bibfield
  {journal} {\bibinfo  {journal} {JHEP}\ }\textbf {\bibinfo {volume} {05}},\
  \bibinfo {pages} {057} (\bibinfo {year} {2021})},\ \Eprint
  {http://arxiv.org/abs/2012.11474} {arXiv:2012.11474 [hep-ph]} \BibitemShut
  {NoStop}%
\bibitem [{\citenamefont {Buonocore}\ \emph
  {et~al.}(2020{\natexlab{a}})\citenamefont {Buonocore}, \citenamefont
  {Haisch}, \citenamefont {Nason}, \citenamefont {Tramontano},\ and\
  \citenamefont {Zanderighi}}]{Buonocore:2020erb}%
  \BibitemOpen
  \bibfield  {author} {\bibinfo {author} {\bibfnamefont {L.}~\bibnamefont
  {Buonocore}}, \bibinfo {author} {\bibfnamefont {U.}~\bibnamefont {Haisch}},
  \bibinfo {author} {\bibfnamefont {P.}~\bibnamefont {Nason}}, \bibinfo
  {author} {\bibfnamefont {F.}~\bibnamefont {Tramontano}}, \ and\ \bibinfo
  {author} {\bibfnamefont {G.}~\bibnamefont {Zanderighi}},\ }\href {\doibase
  10.1103/PhysRevLett.125.231804} {\bibfield  {journal} {\bibinfo  {journal}
  {Phys. Rev. Lett.}\ }\textbf {\bibinfo {volume} {125}},\ \bibinfo {pages}
  {231804} (\bibinfo {year} {2020}{\natexlab{a}})},\ \Eprint
  {http://arxiv.org/abs/2005.06475} {arXiv:2005.06475 [hep-ph]} \BibitemShut
  {NoStop}%
\bibitem [{\citenamefont {Greljo}\ and\ \citenamefont
  {Selimovic}(2021)}]{Greljo:2020tgv}%
  \BibitemOpen
  \bibfield  {author} {\bibinfo {author} {\bibfnamefont {A.}~\bibnamefont
  {Greljo}}\ and\ \bibinfo {author} {\bibfnamefont {N.}~\bibnamefont
  {Selimovic}},\ }\href {\doibase 10.1007/JHEP03(2021)279} {\bibfield
  {journal} {\bibinfo  {journal} {JHEP}\ }\textbf {\bibinfo {volume} {03}},\
  \bibinfo {pages} {279} (\bibinfo {year} {2021})},\ \Eprint
  {http://arxiv.org/abs/2012.02092} {arXiv:2012.02092 [hep-ph]} \BibitemShut
  {NoStop}%
\bibitem [{\citenamefont {Buonocore}\ \emph
  {et~al.}(2020{\natexlab{b}})\citenamefont {Buonocore}, \citenamefont {Nason},
  \citenamefont {Tramontano},\ and\ \citenamefont
  {Zanderighi}}]{Buonocore:2020nai}%
  \BibitemOpen
  \bibfield  {author} {\bibinfo {author} {\bibfnamefont {L.}~\bibnamefont
  {Buonocore}}, \bibinfo {author} {\bibfnamefont {P.}~\bibnamefont {Nason}},
  \bibinfo {author} {\bibfnamefont {F.}~\bibnamefont {Tramontano}}, \ and\
  \bibinfo {author} {\bibfnamefont {G.}~\bibnamefont {Zanderighi}},\ }\href
  {\doibase 10.1007/JHEP08(2020)019} {\bibfield  {journal} {\bibinfo  {journal}
  {JHEP}\ }\textbf {\bibinfo {volume} {08}},\ \bibinfo {pages} {019} (\bibinfo
  {year} {2020}{\natexlab{b}})},\ \Eprint {http://arxiv.org/abs/2005.06477}
  {arXiv:2005.06477 [hep-ph]} \BibitemShut {NoStop}%
\bibitem [{\citenamefont {Buonocore}\ \emph {et~al.}(2022)\citenamefont
  {Buonocore}, \citenamefont {Greljo}, \citenamefont {Krack}, \citenamefont
  {Nason}, \citenamefont {Selimovic}, \citenamefont {Tramontano},\ and\
  \citenamefont {Zanderighi}}]{Buonocore:2022msy}%
  \BibitemOpen
  \bibfield  {author} {\bibinfo {author} {\bibfnamefont {L.}~\bibnamefont
  {Buonocore}}, \bibinfo {author} {\bibfnamefont {A.}~\bibnamefont {Greljo}},
  \bibinfo {author} {\bibfnamefont {P.}~\bibnamefont {Krack}}, \bibinfo
  {author} {\bibfnamefont {P.}~\bibnamefont {Nason}}, \bibinfo {author}
  {\bibfnamefont {N.}~\bibnamefont {Selimovic}}, \bibinfo {author}
  {\bibfnamefont {F.}~\bibnamefont {Tramontano}}, \ and\ \bibinfo {author}
  {\bibfnamefont {G.}~\bibnamefont {Zanderighi}},\ }\href@noop {} {\  (\bibinfo
  {year} {2022})},\ \Eprint {http://arxiv.org/abs/2209.02599} {arXiv:2209.02599
  [hep-ph]} \BibitemShut {NoStop}%
\bibitem [{\citenamefont {Allwicher}\ \emph
  {et~al.}(2022{\natexlab{b}})\citenamefont {Allwicher}, \citenamefont
  {Faroughy}, \citenamefont {Jaffredo}, \citenamefont {Sumensari},\ and\
  \citenamefont {Wilsch}}]{Allwicher:2022mcg}%
  \BibitemOpen
  \bibfield  {author} {\bibinfo {author} {\bibfnamefont {L.}~\bibnamefont
  {Allwicher}}, \bibinfo {author} {\bibfnamefont {D.~A.}\ \bibnamefont
  {Faroughy}}, \bibinfo {author} {\bibfnamefont {F.}~\bibnamefont {Jaffredo}},
  \bibinfo {author} {\bibfnamefont {O.}~\bibnamefont {Sumensari}}, \ and\
  \bibinfo {author} {\bibfnamefont {F.}~\bibnamefont {Wilsch}},\ }\href@noop {}
  {\  (\bibinfo {year} {2022}{\natexlab{b}})},\ \Eprint
  {http://arxiv.org/abs/2207.10756} {arXiv:2207.10756 [hep-ph]} \BibitemShut
  {NoStop}%
\bibitem [{\citenamefont {Feruglio}\ \emph {et~al.}(2017)\citenamefont
  {Feruglio}, \citenamefont {Paradisi},\ and\ \citenamefont
  {Pattori}}]{Feruglio:2017rjo}%
  \BibitemOpen
  \bibfield  {author} {\bibinfo {author} {\bibfnamefont {F.}~\bibnamefont
  {Feruglio}}, \bibinfo {author} {\bibfnamefont {P.}~\bibnamefont {Paradisi}},
  \ and\ \bibinfo {author} {\bibfnamefont {A.}~\bibnamefont {Pattori}},\ }\href
  {\doibase 10.1007/JHEP09(2017)061} {\bibfield  {journal} {\bibinfo  {journal}
  {JHEP}\ }\textbf {\bibinfo {volume} {09}},\ \bibinfo {pages} {061} (\bibinfo
  {year} {2017})},\ \Eprint {http://arxiv.org/abs/1705.00929} {arXiv:1705.00929
  [hep-ph]} \BibitemShut {NoStop}%
\bibitem [{\citenamefont {Allwicher}\ \emph
  {et~al.}(2022{\natexlab{c}})\citenamefont {Allwicher}, \citenamefont
  {Isidori},\ and\ \citenamefont {Selimovic}}]{Allwicher:2021ndi}%
  \BibitemOpen
  \bibfield  {author} {\bibinfo {author} {\bibfnamefont {L.}~\bibnamefont
  {Allwicher}}, \bibinfo {author} {\bibfnamefont {G.}~\bibnamefont {Isidori}},
  \ and\ \bibinfo {author} {\bibfnamefont {N.}~\bibnamefont {Selimovic}},\
  }\href {\doibase 10.1016/j.physletb.2022.136903} {\bibfield  {journal}
  {\bibinfo  {journal} {Phys. Lett. B}\ }\textbf {\bibinfo {volume} {826}},\
  \bibinfo {pages} {136903} (\bibinfo {year} {2022}{\natexlab{c}})},\ \Eprint
  {http://arxiv.org/abs/2109.03833} {arXiv:2109.03833 [hep-ph]} \BibitemShut
  {NoStop}%
\bibitem [{\citenamefont {Altmannshofer}\ \emph {et~al.}(2019)\citenamefont
  {Altmannshofer} \emph {et~al.}}]{Belle-II:2018jsg}%
  \BibitemOpen
  \bibfield  {author} {\bibinfo {author} {\bibfnamefont {W.}~\bibnamefont
  {Altmannshofer}} \emph {et~al.} (\bibinfo {collaboration} {Belle-II}),\
  }\href {\doibase 10.1093/ptep/ptz106} {\bibfield  {journal} {\bibinfo
  {journal} {PTEP}\ }\textbf {\bibinfo {volume} {2019}},\ \bibinfo {pages}
  {123C01} (\bibinfo {year} {2019})},\ \bibinfo {note} {[Erratum: PTEP 2020,
  029201 (2020)]},\ \Eprint {http://arxiv.org/abs/1808.10567} {arXiv:1808.10567
  [hep-ex]} \BibitemShut {NoStop}%
\bibitem [{\citenamefont {Iguro}\ \emph {et~al.}(2022)\citenamefont {Iguro},
  \citenamefont {Kitahara},\ and\ \citenamefont {Watanabe}}]{Iguro:2022yzr}%
  \BibitemOpen
  \bibfield  {author} {\bibinfo {author} {\bibfnamefont {S.}~\bibnamefont
  {Iguro}}, \bibinfo {author} {\bibfnamefont {T.}~\bibnamefont {Kitahara}}, \
  and\ \bibinfo {author} {\bibfnamefont {R.}~\bibnamefont {Watanabe}},\
  }\href@noop {} {\  (\bibinfo {year} {2022})},\ \Eprint
  {http://arxiv.org/abs/2210.10751} {arXiv:2210.10751 [hep-ph]} \BibitemShut
  {NoStop}%
\end{thebibliography}%
\end{document}